\RequirePackage{tikz}

\documentclass[pdflatex, sn-basic]{sn-jnl}

\date{}

\jyear{2021}%

\usepackage[utf8]{inputenc}

\usepackage{graphicx}
\usepackage{amsmath}
\usepackage{amsthm}
\usepackage{amsfonts}
\usepackage{amssymb}
\usepackage{comment}

\usepackage{mathrsfs}
\usepackage{calligra}
\usepackage{mathtools}

\usepackage{pgfplots, pgfplotstable}
\pgfplotsset{width=10cm, compat=1.9}

\usepackage{float}

\theoremstyle{definition}

\begin{document}
	
	\title[A sparse matrix formulation of model-based ensemble Kalman
	filter]{A sparse matrix formulation of model-based ensemble Kalman
		filter}
	
	
	\author*{\fnm{Håkon} \sur{Gryvill}}\email{hakon.gryvill@ntnu.no}
	
	\author{\fnm{Håkon} \sur{Tjelmeland}}\email{haakon.tjelmeland@ntnu.no}
	
	\affil{\orgdiv{Department of Mathematical Sciences}, \orgname{Norwegian University of Science and Technology}, \orgaddress{\street{Alfred Getz' vei 1}, \city{Trondheim}, \postcode{7034}, \country{Norway}}}

	
	\abstract{We introduce a computationally efficient variant of the model-based ensemble Kalman filter (EnKF). We propose two changes to the original formulation. First, we phrase the setup in terms of precision matrices instead of covariance matrices, and introduce a new prior for the precision matrix which ensures it to be sparse. Second, we propose to split the state vector into several blocks and formulate an approximate updating procedure for each of these blocks. 
		
		We study in a simulation example the computational speedup and the approximation error resulting from using the proposed approach. The speedup is substantial for high dimensional state vectors, allowing the proposed filter to be run on much larger problems than can be done with the original formulation. In the simulation example the approximation error resulting from using the introduced block updating is negligible compared to the Monte Carlo variability inherent in both the original and the proposed procedures.
	}
	
	
	\maketitle

\section{Introduction}\label{section:intro}

State-space models are frequently used in many applications. Examples of application areas are economics \citep{Creal2011,ChanStrachan2020}, weather forecasting  \citep{HoutekamerZhang2016,HottaOta2021}, signal processing \citep{LoeligerEtAl2007} and neuroscience \citep{SmithBrownEmery2003}. A state-space model consists of an unobserved latent $\{x_t\}$ discrete time Markov process and a related observed $\{y_t\}$ process, where $y_t$ gives information about  $x_t$, and the $y_t$'s are assumed to be conditionally independent given the $\{x_t\}$ process. Given observed values $y_1,\ldots,y_t$ the goal is to do inference about one or more of the $x_t$'s. In this article the focus is on the filtering problem, where the goal is to find the conditional distribution of $x_t$ given observations $y_{1:t}=(y_1,\ldots,y_t)$. The filtering problem is also known as sequential data assimilation and online inference.

The Markov structure in the specification of the state-space model allows the filtering problem to be solved recursively. Having available a solution of the filtering problem at time $t-1$, i.e. the distribution $p(x_{t-1}\vert y_{1:t-1})$, one can first use this filtering distribution to find the one-step forecast distribution $p(x_t\vert y_{1:t-1})$, which one in turn can combine with the observed data at the next time step, $y_t$, to find $p(x_t\vert y_{1:t})$. The identification of the forecast distribution is often termed the forecast or prediction step, whereas the process of finding the filtering distribution is called the update or analysis step. If a linear-Gaussian model is assumed analytical solutions are available both for the forecast and update steps, and is known as the Kalman filter \citep{Kalman1960,KalmanBucy1961}. Another situation where an analytical solution is available for the forecast and update steps is when $x_t$ is a vector of categorical or discrete variables. In essentially all other situations, however, analytical solutions are not available and one has to resort to approximate solutions. Ensemble-based methods are here the most popular choice, where each distribution of interest is represented by a set of particles, or an ensemble of realisations, from the distribution of interest. In ensemble-based methods one also alternates between a forecast and an update step. The forecast step is straightforward to implement without approximations in ensemble methods, whereas the update step is challenging. In the update step an ensemble of realisations, $x_t^{(1)},\ldots, x_t^{({\mathcal M})}$, from the forecast distribution $p(x_t\vert y_{1:t-1})$ is available and the task is to update each realisation $x_t^{(m)}$ to a corresponding realisation $\widetilde{x}_t^{(m)}$ from the filtering distribution $p(x_t\vert y_{1:t})$. The forecast distribution $p(x_t\vert y_{1:t-1})$ serves as a prior in the update step, and the filtering distribution $p(x_t\vert y_{1:t})$ is the resulting posterior. In the following we therefore refer to the ensemble of realisations from the forecast distribution as the prior ensemble and the ensemble of realisations from the filtering distribution as the posterior ensemble.

There are two popular classes of ensemble filtering methods, particle filters \citep{DoucetEtAl2001,DoucetJohansen2011} and ensemble Kalman filters (EnKFs) \citep{BurgersEtAl1998,Evensen2009}. Particle filters are based on importance sampling and resampling and has the advantage that it can be shown to converge to the correct filtering solution in the limit when the number of particles goes towards infinity.  However, in applications with high dimensional state vectors, $x_t$, particle filters typically collapse when run with a finite number of particles. The updating procedure in EnKFs is based on a linear-Gaussian model, but experience in applications is that EnKFs  produce good results also in situations where the linear-Gaussian assumptions are not fulfilled. However, except when the linear-Gaussian model is correct the EnKF is only approximate, even in the limit when the number of ensemble elements goes towards infinity. Most of the EnKF literature is in applied journals and has not much focus on the underlying statistical theory, but some publications has also appeared in statistical journals \citep{SaetromOmre2013,Katzfuss,Katzfuss2020, LoeTjelmeland2020, LoeTjelmeland2022}.

In EnKF the update step consists of two parts. First a prior covariance matrix is established based on the prior ensemble, which thereafter is used to modify each prior ensemble element to a corresponding posterior element. Many variants of the original EnKF algorithm of \citet{EnKF} have been proposed and studied in the literature, see for example \citet{Evensen2009} and \citet{HoutekamerZhang2016} and references therein. In particular the original algorithm of \citet{EnKF} is known to severely underestimate the uncertainty and much attention has focused on how to correct for this. The most frequent approach to this problem is the rather ad hoc solution of variance inflation, see the discussion in \citet{LuoHoteit1997}. \citet{HoutekamerMitchell1997} identified the source of the problem to be inbreeding, that the same prior ensemble elements that are first used to establish the prior covariance matrix are thereafter modified into posterior ensemble elements. To solve this problem \citet{HoutekamerMitchell1997} proposed to split the prior ensemble in two, where the prior ensemble elements in one part is used to establish a prior covariance matrix which is used when modifying the prior ensemble elements in the other part into corresponding posterior elements. As discussed in \citet{HoutekamerZhang2016} this cross validation setup is later generalised to a situation where the prior ensemble is split into more than two parts. \citet{MyrsethEtAl2013} propose to use a resampling approach to solve the inbreeding problem. Another issue with the original EnKF setup of \citet{Evensen2009} that has been focused in the literature is that the empirical covariance matrix of the prior ensemble is used as the prior covariance matrix, thereby effectively ignoring the associated estimation uncertainty. To cope with this problem \citet{HEnKF} set the problem of identifying the prior covariance matrix in a Bayesian setting. {To obtain an analytically tractable posterior distribution for the mean vector and the covariance matrix of the prior ensemble, the natural conjugate normal inverse Wishart prior is adopted. For each ensemble element, a sample from the resulting normal inverse Wishart posterior distribution is generated and used in the updating of that ensemble element.} 
Corresponding Bayesian schemes for establishing the prior covariance matrix are later adopted in \citet{Bocquet2011}, \citet{BocquetEtAl2015} and \citet{TsyrulnikovRakitko2017}. \citet{LoeTjelmeland2021} propose to base the updating step of the EnKF on an assumed Bayesian model for all variables involved. The updating step is defined by restricting it to be consistent with the assumed model at the same time as it should be robust against modeling error. Both the need for defining a prior for the covariance matrix and the cross validation approach discussed above comes as necessary consequences of this setup. In addition it entails that the new data $y_t$ should be taken into account when generating the prior covariance matrix.

In the present article we adopt the model-based EnKF approach of \citet{LoeTjelmeland2021}, but our focus is to formulate a computationally efficient variant of this approach. To obtain this we do two important changes. First, in \citet{LoeTjelmeland2021} {the natural conjugate} normal inverse Wishart prior is used for the mean vector and covariance matrix of the prior ensemble elements. When sampling from the corresponding posterior distribution this results in a covariance matrix that is full. To update a prior ensemble element to the corresponding posterior element the covariance matrices must be used in a series of matrix operations, such as various matrix decompositions and multiplications with vectors. We rephrase the approach of \citet{LoeTjelmeland2021} to use precision matrices, or inverse covariance matrices, instead of covariance matrices and propose {to use a Gaussian partially ordered Markov model \citep{CressieDavidson1998} to construct a prior for the mean vector and the precision matrix of the prior ensemble. This prior is also conjugate for the assumed normal likelihood for the ensemble elements, and the resulting posterior is therefore analytically tractable. Moreover, with this prior distribution we are able to ensure that the precision matrices sampled from the
resulting posterior distribution are band matrices.}
With a band structure in the precision matrix some of the matrix operations in the updating of a prior ensemble element to the corresponding posterior element can be done more efficiently. However, some of the matrix operations, singular value decompositions, do not benefit from this band structure. To make also this part of the procedure computationally more efficient we propose to do an approximation which allows us to replace the singular value decomposition of one large matrix with singular value decompositions of several smaller matrices.

{Compared to the updating procedure in the original EnKF setup of 
\citet{EnKF}, the updating procedure resulting from adopting the Bayesian scheme introduced in
\citet{HEnKF} is clearly computationally much more expensive. The increased computational demands come for two reasons. First, the Bayesian setup requires one covariance matrix to be generated for each ensemble element, whereas the original EnKF setup uses the same covariance matrix for all ensemble elements. Second, the covariance matrices generated by the original Bayesian setup of \citet{HEnKF} are full and without any special structure, whereas the covariance matrix used in the original EnKF setup has a low rank which can be used to speed up the necessary matrix computations. In the present paper we resolve the second issue by adopting a prior model producing sparse precision matrices. Regarding the first issue, \citet{LoeTjelmeland2021} demonstrated that with their model-based EnKF approach the generated ensemble gave a realistic representation of the uncertainty, whereas the original EnKF setup produced ensembles which underestimated the uncertainty. For small ensemble sizes the underestimation of the uncertainty when adopting the original EnKF setup is severe, whereas with large enough ensemble sizes it becomes negligible. To obtain a realistic representation of the uncertainty we therefore have two possible strategies. One can either adopt the model-based EnKF approach used in \citet{LoeTjelmeland2021} and in the present paper, or one can use the original EnKF setup with a very large number of ensemble elements. Which of the two approaches that are computationally more efficient depends on the computational cost of running the forward function. If the forward function is computationally expensive, so that this part dominates the total computing time of the filter, it is clearly computationally most efficient to adopt the model-based EnKF procedure introduced in the present paper. If the computation time is not dominated by the running of the forward function, the picture is less clear. Depending on how many ensemble elements that are necessary to get a sufficiently small underestimation of the uncertainty, the original EnKF with a large number of ensemble elements may be the computationally most efficient alternative.}

This article is structured as follows. In Section \ref{section:preliminaries} we first review some numerical algorithms for sparse matrices and some properties of the Gaussian density. In the same section we also present the state-space model. In Section \ref{section:bayes_upd_ensemble} we rephrase the approach introduced in \citet{LoeTjelmeland2021} to use precision matrices instead of covariance matrices. The new prior for the precision matrix is proposed in Section \ref{section:remaining_upd_elems}. In Section \ref{section:block_upd} we propose a computationally efficient approximation of the updating procedure presented in Section \ref{section:bayes_upd_ensemble}. We present results {of simulation examples} in Section \ref{section:sim_examples}, and finally we provide some closing remarks in Section \ref{section:closing_remarks}.


\section{Preliminaries}\label{section:preliminaries}

This section introduces material necessary to understand the approach proposed in later sections. We start by discussing some numerical properties of sparse matrices, thereafter review how the multivariate Gaussian distribution can be formulated in terms of precision matrices. Lastly, we introduce the state-space model and provide a brief introduction on simulation-based techniques.


\subsection{Numerical properties of sparse matrices}\label{section:num_prop_sparse_matrix}




Suppose that $x, y \in \mathbb{R}^n$, that $Q \in \mathbb{R}^{n\times n}$ is a given symmetric positive definite band matrix with bandwidth $p$, where $n \gg p$, that $y$ is given, and that we want to solve the matrix equation 
\begin{equation}\label{chol_example}
    Qx = y
\end{equation}
with respect to $x$. Since we assume that $Q$ is a symmetric matrix, we can solve the equation above using the Cholesky decomposition $Q=L L^T$, where $L \in \mathbb{R}^{n\times n}$ is a lower triangular matrix. Due to the band structure of $Q$ we know that $L$ has lower bandwidth $p$, which in turn enables us to compute $L$ with complexity $\mathcal{O}(np^2)$, see Algorithm 2.9 in \citet{GMRF}. Since we now are able to compute $L$ efficiently, we can make use of Algorithm 2.1 in \citet{GMRF} to solve \eqref{chol_example} efficiently as well. In the following, we describe the algorithm step by step. 

First, we rewrite \eqref{chol_example} as
\begin{equation} \label{chol_example_LLT}
    LL^T x = y.
\end{equation}
If we define $L^T x = v$, we can solve \eqref{chol_example} by first solving $Lv = y$ for $v$ and then solving $L^T x = v$ for $x$. We can exploit the band structure of $L$ to solve these equations efficiently. Using that $L$ is lower triangular, we can solve $Lv = y$ row by row, using "forward substitution" \citep[p.~32]{GMRF}. We denote the $(i,j)$th entry of $L$ as $L^{i,j}$ and the $i$th element of $y$ as $y^i$. The $i$th element of $v$, denoted $v^i$, is computed as follows 
\begin{equation}\label{cholesky_forward_substitution}
    v^i = \frac{1}{L^{i,i}}(y^i-\sum_{j=\max\{0, i-p\}}^{i-1} L^{i,j}v^j).
\end{equation}
Similarly, we can solve $v = L^T x$ for $x$ using "backward substitution"
\begin{equation}\label{cholesky_back_substitution}
    x^i = \frac{1}{L^{i,i}}(v^i-\sum_{j=i+1}^{\min\{i+p,n\}} L^{j,i}x^j).
\end{equation}
Notice that we compute the entries of $x$ "backwards"; we first compute $x^n$ and then move our way backwards to $x^1$. If we again assume $n \gg p$, the computational complexity of forward and backward substitution is $\mathcal{O}(np)$. This means that the overall complexity of computing $x$ using the presented approach is $\mathcal{O}(np^2)$. Note that solving \eqref{chol_example} with the "brute force" approach, i.e. computing $x=Q^{-1}y$, has computational complexity $\mathcal{O}(n^3)$.

Backward substitution can also be used to sample efficiently from a Gaussian distribution when the precision matrix is a band matrix. Assume that $z$ is a vector of standard normally distributed variables and that we want to simulate $x$ from a Gaussian distribution with mean $0$ and covariance matrix $Q^{-1}$. We can simulate $x$ by solving
\begin{equation}
    L^T x = z
\end{equation}
using backward substitution, as specified in \eqref{cholesky_back_substitution}. When $Q$ is a band matrix with bandwidth $p$, where $n \gg p$, the computational complexity is $\mathcal{O}(np^2)$. However, when $Q$ is full the computational complexity of simulating $x$ is $\mathcal{O}(n^3)$.

\subsection{Gaussian distribution phrased with precision matrices} \label{section:norm_dist_prec_matr}


Let $x \in \mathbb{R}^n$ have a multivariate Gaussian distribution with mean $\mu \in \mathbb{R}^n$ and precision matrix $Q \in \mathbb{R}^{n\times n}$. We let ${\mathcal N}(x; \mu, Q)$ denote the density function of this distribution, i.e. 
\begin{equation} \label{normal_pdf_prec_mx}
    {\mathcal N}(x; \mu, Q) = (2\pi)^{-n/2} \vert Q\vert ^{1/2}\exp{\left(-\frac{1}{2}(x-\mu)^T Q (x-\mu) \right)}.
\end{equation}
Defining $A \subset \{1, \dots n\}$ and $B=\{1, \dots , n\}\setminus A$, we can partition $x$, $\mu$ and $Q$ into blocks
\begin{equation}
    x = 
    \begin{pmatrix}
    x^A \\ x^B
    \end{pmatrix},  \quad
    \mu = 
    \begin{pmatrix}
    \mu^A \\ \mu^B
    \end{pmatrix},  \quad
    Q = 
    \begin{pmatrix}
    Q^{AA} & Q^{AB} \\ Q^{BA} & Q^{BB}
    \end{pmatrix}.
\end{equation}
According to Theorem 2.5 in \citet{GMRF}, we then have that 
\begin{equation} \label{cond_dist_norm_prec}
    p(x^A\vert x^B) = {\mathcal N}(x^A; \mu^A - (Q^{AA})^{-1}Q^{AB}(x^B-\mu^B), Q^{AA}).
\end{equation}
Similarly,
\begin{equation}
    p(x^B) = {\mathcal N}(x^B; \mu^B, Q^{BB}-(Q^{AB})^T (Q^{AA})^{-1}Q^{AB}).
\end{equation}
The last expression can be derived by picking out the parts of the mean vector $\mu$ and covariance matrix $Q^{-1}$ that corresponds to $x^B$. When we have found the covariance matrix of $x^B$ we find the precision matrix by inversion. From the first expression we see that computing the precision matrix for $x^A\vert x^B$ is particularly easy when the Gaussian distribution is formulated with precision matrices. 

\subsection{State-space model}\label{section:SSM}

\begin{figure}
    \begin{center}
        \begin{tikzpicture}[scale=0.9]
        \draw[black,thick] (0,0) circle (0.40cm);
        \draw[black,thick] (0,0) node {$x_1$};
        
        \draw[black,thick] (2,0) circle (0.40cm);
        \draw[black,thick] (2,0) node {$x_2$};
        
        \draw[black,thick,->] (0.4,0) -- (1.6,0);
        \draw[black,thick,->] (2.4,0) -- (3,0);
        
        \draw[black,thick,->] (4,0) -- (4.6,0);
        \draw[black,thick,dashed] (3,0) -- (4,0);

        \draw[black,thick] (5,0) circle (0.40cm);
        \draw[black,thick] (5,0) node {$x_{t-1}$};
        
        \draw[black,thick] (7,0) circle (0.40cm);
        \draw[black,thick] (7,0) node {$x_t$};
        
        \draw[black,thick,->] (5.4,0) -- (6.6,0);
        \draw[black,thick,->] (7.4,0) -- (8,0);
        
        \draw[black,thick,->] (9,0) -- (9.6,0);
        \draw[black,thick,dashed] (8,0) -- (9,0);
        
        \draw[black,thick] (10,0) circle (0.40cm);
        \draw[black,thick] (10,0) node {$x_T$};

        \draw[black,thick,->] (0,0.4) -- (0,1);
        \draw[black,thick,->] (2,0.4) -- (2,1);
        \draw[black,thick,->] (5,0.4) -- (5,1);
        \draw[black,thick,->] (7,0.4) -- (7,1);
        \draw[black,thick,->] (10,0.4) -- (10,1);
        
        \draw[black,thick] (0,1.4) circle (0.40cm);
        \draw[black,thick] (0,1.4) node {$y_1$};
        \draw[black,thick] (2,1.4) circle (0.40cm);
        \draw[black,thick] (2,1.4) node {$y_2$};
        \draw[black,thick] (5,1.4) circle (0.40cm);
        \draw[black,thick] (5,1.4) node {$y_{t-1}$};
        \draw[black,thick] (7,1.4) circle (0.40cm);
        \draw[black,thick] (7,1.4) node {$y_t$};
        \draw[black,thick] (10,1.4) circle (0.40cm);
        \draw[black,thick] (10,1.4) node {$y_T$};
        \end{tikzpicture}

    \end{center}

\caption{DAG illustration of a state-space model. The edges illustrate the stochastic dependencies between the nodes. The latent variables $x_t, t = 1, \dots, T$ are unobserved, while $y_t, t = 1, \dots, T$ are observations}
\label{fig:StateSpace}
\end{figure}
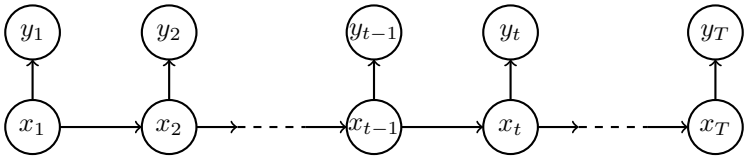

\begin{sloppypar}
A state-space model \citep{ShumwayStoffer, BrockwellDavis} consists of a set of latent variables, denoted ${\{x_t\}_{t=1}^T}$, ${x_t \in \mathbb{R}^{n_x}}$, and a set of observations $\{y_t\}_{t=1}^T$, ${y_t \in \mathbb{R}^{n_y}}$. The latent variables follow a first order Markov chain with initial distribution $p(x_1)$ and transition probabilities $p(x_t\vert x_{t-1})$, $t\geq 2$. That is, the joint distribution for $x_{1:T} = (x_1, \dots , x_T)$ can be written as
\end{sloppypar}

\begin{equation}
    p(x_{1:T}) = p(x_1) \prod_{t=2}^T p(x_t\vert x_{t-1}).
\end{equation}
In addition, each observation $y_t$ is considered to be conditionally independent of the remaining observations, given $x_t$. The joint likelihood for the observations $y_{1:T}=(y_1, \dots , y_T)$ can be formulated as
\begin{equation}
    p(y_{1:T}\vert x_{1:T}) = \prod_{t=1}^T p(y_t \vert x_t).
\end{equation}
A state-space model is illustrated by a directed acyclic graph (DAG) in Figure \ref{fig:StateSpace}, where each variable is represented with a node and the edges symbolise the dependencies between the variables.

The main interest of this article, and the reason we introduce the state-space model, is to assess the filtering problem. The objective of the filtering problem is to compute the filtering distribution, $p(x_t\vert y_{1:t})$, for $t = 1, \dots , T$. That is, we want to find the distribution for the latent variable at time $t$, i.e. $x_t$, given all of the observations up to the same time step, $y_{1:t}$. Due to the Markov assumptions made in the state-space model, we are in principle able to assess this quantity sequentially. Each iteration is performed in two steps, namely
\begin{align}
    p(x_t\vert y_{1:t-1}) &= \int
    p(x_t\vert x_{t-1})p(x_{t-1}\vert y_{1:t-1})dx_{t-1},\label{forecast_exp}\\
    p(x_t\vert y_{1:t}) &= \frac{p(x_t\vert y_{1:t-1})p(y_t\vert x_t)}{\int p(x_t\vert y_{1:t-1})p(y_t\vert x_t)dx_t}\label{filter_exp}.
\end{align}
The first expression is the prediction step and gives the forecast distribution $p(x_t\vert y_{1:t-1})$, while the second expression is called the update step and yields the filtering distribution $p(x_t\vert y_{1:t})$. Note that the filtering distribution is computed through Bayes' rule, where $p(x_t \vert  y_{1:t-1})$ is the prior, $p(y_t\vert x_t)$ is the likelihood and $p(x_t\vert y_{1:t})$ is the posterior. In the following we therefore use the terms prior and forecast, and the terms posterior and filtering, interchangeably. In this article, our main focus is on the update step. 

When the state-space model is linear and Gaussian, the expressions can be solved analytically through the Kalman filter \citep{Kalman1960}. However, the prediction and update steps in \eqref{forecast_exp} and \eqref{filter_exp} are generally speaking not feasible to solve analytically. Hence approximative methods are used. A common approach is to use simulation-based techniques, where a set of realisations, which is called an ensemble, are used to explore the state-space model by moving the realisations forward in time according to the state-space model. The filtering and forecast distributions are then represented by an ensemble at each time step, which is initially sampled from $p(x_1)$. The following paragraph provides an overview of how this is done for one iteration.

Assume that a set of ${\mathcal M}$ independent realisations $\{\widetilde{x}_{t-1}^{(1)}, \dots , \widetilde{x}_{t-1}^{({\mathcal M})}\}$ from $p(x_{t-1}\vert y_{1:t-1})$ is available at time $t$. If we are able to simulate from the forward model $p(x_t\vert x_{t-1})$, we can obtain ${\mathcal M}$ independent realisations from $p(x_t\vert y_{1:t-1})$ by simulating from $x_t^{(m)}\vert \widetilde{x}_{t-1}^{(m)} \sim p(x_t\vert \widetilde{x}_{t-1}^{(m)})$ independently for $m = 1, \dots , {\mathcal M}$. This is the prediction step, and can usually be performed without any approximations. Next, we use the prediction, or prior, ensemble $\{{x}_{t-1}^{(1)}, \dots , {x}_{t-1}^{({\mathcal M})}\}$ to obtain samples from the filtering distribution $p(x_t\vert y_{1:t})$, which is often called a posterior ensemble. This can be done by conditioning the samples from the forecast distribution, $\{x_t^{(1)}, \dots , x_t^{({\mathcal M})}\}$, on the new observation $y_t$. This step is called the update step and is generally not analytically feasible. Hence approximative methods are necessary. In the following section, we present a procedure that enables us to carry out the update step. 

\section{Model-based EnKF} \label{section:bayes_upd_ensemble} 

In this section we start by reviewing the model-based EnKF framework introduced in
\citet{LoeTjelmeland2021}. The focus in our presentation is on the
underlying model framework, the criterion used for selecting the particular chosen update,
and on the resulting updating procedure. We do not include the mathematical derivations
leading to the computational procedure. Moreover, we phrase the framework in terms of
precision matrices, whereas \citet{LoeTjelmeland2021} use covariance matrices.

The focus in the section is on how to use a prior ensemble $\{x_t^{(1)},\ldots,x_t^{({\mathcal M})}\}$ to update one of these prior ensemble elements, number $m$ say, to a corresponding posterior ensemble element $\widetilde{x}_t^{(m)}$. All the variables involved in this operation are associated with the same time
$t$. To simplify the notation we therefore omit the time subscript $t$ in the following discussion. So we write $\{x^{(1)},\ldots,x^{({\mathcal M})}\}$ instead of $\{x_t^{(1)},\ldots,x_t^{({\mathcal M})}\}$ for the available prior ensemble, we write $\widetilde{x}^{(m)}$ instead of $\widetilde{x}_t^{(m)}$ for the generated posterior ensemble element number $m$, we write $x$ instead of $x_t$ for the latent variable at time $t$,
and we write $y$ instead of $y_t$ for the new data that becomes available at time $t$.

\subsection{Assumed Bayesian model}\label{section:bayesian_model}

\begin{figure}
    \begin{center}
        \begin{tikzpicture}[scale=0.9]
        
        \draw[black, thick] (3,0) circle (0.50cm);
        \draw[black, thick] (3,0) node {$x^{(m)}$};

        \draw[black, thick] (7,0) circle (0.50cm);
        \draw[black, thick] (7,0) node {$x$};

        \draw[black, thick] (3,3) circle (0.50cm);
        \draw[black, thick] (3,3) node {$\widetilde{x}^{(m)}$};

        \draw[black, thick] (7,3) circle (0.50cm);
        \draw[black, thick] (7,3) node {$y$};

        \draw[black,thick] (3,-3) circle (0.50cm);
        \draw[black,thick] (3, -3) node {$\theta$};

        \draw[black,thick,->] (3, -2.5) -- (3,-0.5);
        \draw[black,thick,->] (3.35,-2.65) -- (4.75,-0.44);
        \draw[black,thick,->] (2.65,-2.65) -- (1.25,-0.44);
        \draw[black, thick,->] (3.5,-3) -- (6.65, -0.35);
        
        \draw[black,thick,->] (3, 0.5) -- (3,2.5);
    
        \draw[black,thick,->] (7, 0.5) -- (7,2.5);
        \draw[black,thick,->] (6.5,3) -- (3.5, 3);

        \draw[black, thick, ->] (2.75,-2.55) arc (197:162.5:8.6);

        \draw[black, thick] (1,0) circle (0.50cm);
        \draw[black, thick] (1,0) node {$x^{(1)}$};

        \draw[black, thick] (5,0) circle (0.50cm);
        \draw[black, thick] (5,0) node {$x^{({\mathcal M})}$};

        \draw[black,thick,loosely dotted] (1.75,0) -- (2.25,0);
        \draw[black,thick,loosely dotted] (3.75,0) -- (4.25,0);

        \end{tikzpicture}

    \end{center}
    \caption{DAG representation of the assumed Bayesian model for updating the $m$th realisation}
    \label{fig:bayesian_model}

\end{figure}
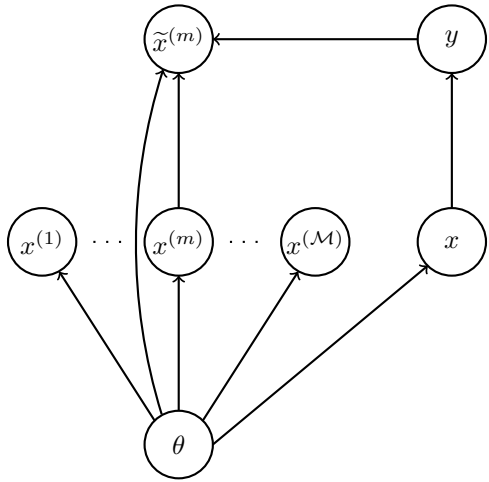

In model-based EnKF the updating of a prior ensemble element $x^{(m)}$ to the corresponding posterior
ensemble element $\widetilde{x}^{(m)}$ is based on an assumed model. The dependence structure of
the assumed model is illustrated in the DAG in Figure \ref{fig:bayesian_model}.
It should be noted that the assumed model is not supposed to be correct, it is just used as a mean to 
formulate a reasonable and consistent
updating procedure. To stress this we follow the notation in \citet{LoeTjelmeland2021} and use $f$
as the generic symbol for all densities associated to the assumed model, whereas we continue to use
$p$ to denote the corresponding correct (and typically unknown) densities. We use subscripts on $f$ to
specify what stochastic variables the density relates to. For example $f_{y\vert x}(y\vert x)$ is the conditional
density for the new observations $y$ given the latent variable $x$.

In the assumed model we let the prior ensemble elements $x^{(1)},\ldots,x^{({\mathcal M})}$ and
the unobserved latent variable $x$ at time $t$ be conditionally independent and identically Gaussian
distributed with a mean vector $\mu$ and a precision matrix $Q$, i.e.
\begin{align}
    f_{x\vert \theta}(x\vert \theta) &= {\cal N}(x; \mu, Q)\label{prior_model},\\ \label{eq:prior}
    f_{x^{(i)}\vert \theta}(x^{(i)}\vert \theta) & = {\cal N}(x^{(i)}; \mu, Q), \quad i = 1, \dots , {\mathcal M},
\end{align}
where $\theta=(\mu,Q)$. For the parameter $\theta$ of this Gaussian density we assume some prior
distribution $f_\theta(\theta)$. \citet{LoeTjelmeland2021} assume a normal inverse Wishart prior
for $(\mu, Q^{-1})$, which implies that $Q$ is full, whereas we want to adopt a prior which ensures that
$Q$ is a band matrix. In Section \ref{section:new_prior} we detail the prior we are using.
The last element of the assumed model is to let the new data $y$ be conditionally
independent of $x^{(1)},\ldots,x^{({\mathcal M})}$ and $\theta$ given $x$, and
\begin{equation} \label{likelihood_model}
    f_{y\vert x}(y\vert x) = {\cal N}(y; Hx, R).
\end{equation}

\begin{sloppypar}
Based on the assumed model the goal is to construct a procedure for generating an updated version
$\widetilde{x}^{(m)}$ of $x^{(m)}$. One should first generate a $\theta=(\mu,Q)$ and
thereafter $\widetilde{x}^{(m)}$. In addition to being a function of $x^{(m)}$ it is natural to allow 
$\widetilde{x}^{(m)}$ to depend on the generated $\theta$ and the new data $y$, as also indicated
in the DAG in Figure \ref{fig:bayesian_model}. The corresponding conditional distribution for
$\widetilde{x}^{(m)}$ we denote by $q(\widetilde{x}^{(m)}\vert x^{(m)},\theta,y)$. The updated
$\widetilde{x}^{(m)}$ is to be used as an (approximate) sample from $p(x_t\vert y_{1:t})$, which in the
ensemble based setting we are considering here is represented by
$p(x_t\vert x_t^{(1)},\ldots,x_t^{({\mathcal M})},y_t)$. However, as we want to use $x^{(m)}$ as a source
for randomness when generating $\widetilde{x}^{(m)}$, the $x^{(m)}$ must be removed from the conditioning set
so one should instead consider $p(x_t\vert x_t^{(1)},\ldots,x_t^{(m-1)},x_t^{(m+1)},\ldots,x_t^{({\mathcal M})},y_t)$.
Under the assumed model this density is equal to $f_{x\vert z^{(m)},y}(x\vert z^{(m)},y)$ when using the shorthand notation 
${z^{(m)}=(x^{(1)},\ldots,x^{(m-1)},x^{(m+1)},\ldots,x^{({\mathcal M})})}$. Thus, we should construct
$q(\widetilde{x}^{(m)}\vert x^{(m)},\theta,y)$ so that
\begin{equation} \label{condition_bayes_upd}
    f_{\widetilde{x}^{(m)}\vert z^{(m)},y}(x\vert z^{(m)},y) = f_{x\vert z^{(m)},y}(x\vert z^{(m)},y)
\end{equation}
holds for all values of $x$, $z^{(m)}$ and $y$. \citet{LoeTjelmeland2021} show that
\eqref{condition_bayes_upd} is fulfilled if $\widetilde{x}^{(m)}$ is generated by first sampling 
$\theta=(\mu,Q)$ from $f_{\theta,z^{(m)},y}(\theta\vert z^{(m)},y)$ and thereafter $\widetilde{x}^{(m)}$ is
sampled according to a $q(\widetilde{x}^{(m)}\vert x^{(m)},\theta,y)$ defined by
\begin{equation}\label{eq:tildexGeneral}
  \widetilde{x}^{(m)} = B(x^{(m)}-\mu)+\mu + K(y-H\mu)+\widetilde{\epsilon}^{(m)},
\end{equation}
where $\widetilde{\epsilon}^{(m)}$ is independent of everything else and generated from a zero-mean
Gaussian distribution with a covariance matrix $S$, $B$ is a matrix connected to the positive
semidefinite $S$ by the relation
\begin{equation}\label{eq:SB}
BQ^{-1}B^T + S = (I-KH)Q^{-1},
\end{equation}
and $K$ is the Kalman gain matrix
\begin{equation}
K = (Q+H^TRH)^{-1}H^TR.
\end{equation}
We note in passing that by inserting for $K$ in \eqref{eq:SB} and thereafter applying the Woodbury identity
\citep{Sherman_Morrison_Woodbury}, one gets ${(I-KH)Q^{-1}=(Q+H^TRH)^{-1}}$, so \eqref{eq:SB} is
equivalent to
\begin{equation}\label{eq:SB2}
  BQ^{-1}B^T+S = (Q+H^TRH)^{-1}.
\end{equation}

\end{sloppypar}
\subsection{Optimality criterion} \label{section:opt_sol}
When applying the updating equation \eqref{eq:tildexGeneral} we have a freedom in the choice of
the matrices $B$ and $S$. The restrictions are that $S$ must be positive semidefinite and that
$B$ and $S$ should be related as specified by \eqref{eq:SB2}. 

Under the assumed model all choices of $B$ and $S$ fulfilling the given restrictions
are equally good, as they are all generating an $\widetilde{x}^{(m)}$ from the same distribution.
When recognising that the assumed model is wrong, however, all solutions are no longer equally good. So
we should choose a solution which is robust against the assumptions made in the assumed model.
\citet{LoeTjelmeland2021} formulate a robust solution as one where the $x^{(m)}$ is changed as little as possible when forming $\widetilde{x}^{(m)}$, under the condition that $B$ and $S$ satisfy \eqref{eq:SB2}. The intuition is that this should allow non-Gaussian
properties in $x^{(m)}$ to be transferred to $\widetilde{x}^{(m)}$. Mathematically the criterion is formulated
as minimising the expected squared Euclidean distance between $x^{(m)}$ and $\widetilde{x}^{(m)}$. Thus, we should minimise
\begin{equation}\label{eq:criterion}
  \mbox{E}\hspace*{-0.05cm}\left[(x^{(m)}-\widetilde{x}^{(m)})^T (x^{(m)}-\widetilde{x}^{(m)})\right],
\end{equation}
with respect to $B$ and $S$ under the restriction \eqref{eq:SB2}, where
the expectation is with respect to the joint distribution of $x^{(m)}$ and $\widetilde{x}^{(m)}$ under the assumed model. Note that \citet{LoeTjelmeland2021} is considering a slightly more general solution by using a Mahalanobis distance. \citet{LoeTjelmeland2021} show that \eqref{eq:criterion} is minimised
under the specified restrictions when $S=0$ and
\begin{equation}\label{eq:Boptimal}
  B = U\Lambda^{-\frac{1}{2}}PF^TD^{\frac{1}{2}}V^T,
\end{equation}
where $U$ and $V$ are orthogonal matrices and $D$ and $\Lambda$ are diagonal matrices given by the singular value
decompositions
\begin{align}\label{eq:svd1}
  Q &= VDV^T,\\ \label{eq:svd2}
  Q+H^TRH &= U\Lambda U^T,
\end{align}
and $P$ and $F$ are orthogonal matrices given by the singular value decomposition
\begin{equation}\label{eq:svd3}
  Z = \Lambda^{-\frac{1}{2}}U^T VD^{-\frac{1}{2}} = PGF^T.
\end{equation}

\subsection{Resulting updating procedure} \label{section:res_upd_proc}
\begin{sloppypar}
The resulting procedure for updating a prior ensemble $x^{(1)},\ldots,x^{({\mathcal M})}$ to the corresponding posterior ensemble ${\widetilde{x}^{(1)},\ldots,\widetilde{x}^{({\mathcal M})}}$ is summarised by the pseudocode in Algorithm \ref{alg:optimal_update}.
\end{sloppypar}

\begin{algorithm}[tbp]
\begin{enumerate}
    \item[] \hspace*{-0.45cm}Given $x^{(1)},\ldots,x^{({\mathcal M})}$, $y$, $H$ and $R$.
    \item[] \hspace*{-0.45cm}For $m=1,\ldots,{\mathcal M}$:
    \item \hspace*{0.45cm}Sample $\theta = (\mu,Q)$ from $f_{\theta\vert z^{(m)},y}(\theta\vert z^{(m)},y)$.
    \item \hspace*{0.45cm}Do singular value decomposition in \eqref{eq:svd1}: $V D V^T = Q$.\label{alg:svd1}
    \item \hspace*{0.45cm}Do singular value decomposition in \eqref{eq:svd2}: $U \Lambda U^T = Q+H^T R H$.\label{alg:svd2}
    \item \hspace*{0.45cm}Evaluate $Z$ in \eqref{eq:svd3}: $Z=\Lambda^{-\frac{1}{2}}U^T VD^{-\frac{1}{2}}$.
    \item \hspace*{0.45cm}Do singular value decomposition in \eqref{eq:svd3}: $PGF^T=Z$.\label{alg:svd3}
    \item \hspace*{0.45cm}Evaluate $B$ in \eqref{eq:Boptimal}: $B = U\Lambda^{-\frac{1}{2}}PF^TD^{\frac{1}{2}}V^T$.\label{alg:B}
    \item \hspace*{0.45cm}Evaluate $\widetilde{x}^{(m)}$: $\widetilde{x}^{(m)} = B(x^{(m)}-\mu)+\mu+K(y-H\mu)$ \label{alg:update_x^m}
\end{enumerate}
\caption{Summary of the resulting updating procedure for computing the posterior ensemble
  $\widetilde{x}^{(1)},\ldots,\widetilde{x}^{({\mathcal M})}$ from the prior ensemble $x^{(1)},\ldots,x^{({\mathcal M})}$.}
\label{alg:optimal_update}
\end{algorithm}

In the following sections our focus is on how to make this updating procedure computationally efficient when
the dimensions of the state vector and corresponding observation vector are large. First, in Section
\ref{section:remaining_upd_elems}, we propose
a new prior for $\theta$ which enables efficient sampling of $\theta=(\mu,Q)$ from the corresponding
posterior $f_{\theta\vert z^{(m)},y}(\theta\vert z^{(m)},y)$. The generated $Q$ is then a sparse matrix, which also limits the
memory usage necessary to store it, since we of course only need to store the non-zero elements. However,
sparsity of $Q$ does not influence the computational efficiency of the singular value decompositions in
Steps \ref{alg:svd1}, \ref{alg:svd2} and \ref{alg:svd3} of Algorithm \ref{alg:optimal_update}. One should
note that we may rephrase Steps \ref{alg:svd1} and \ref{alg:svd2} to use Cholesky decompositions instead, since $Q$ and $Q+H^TRH$ are symmetric and positive definite matrices. Under reasonable assumptions also $Q+H^TRH$ is a sparse matrix so thereby these two steps can be done computationally efficient. 
The $Z$ in Step \ref{alg:svd3}, however, is in general neither a symmetric positive definite matrix, nor 
sparse. To introduce Cholesky decompositions in Steps \ref{alg:svd1} and \ref{alg:svd2} will therefore
not change the order of the computational complexity of the procedure. Instead of substituting the singular value
decompositions in Steps \ref{alg:svd1} and \ref{alg:svd2} with Cholesky decompositions, we therefore
in Section \ref{section:block_upd} propose an approximation of Steps \ref{alg:svd1} to \ref{alg:B}
by splitting the state vector into a series of
blocks and running Steps \ref{alg:svd1} to \ref{alg:B} for each of the blocks separately.

\section{Prior model and sampling of \texorpdfstring{$\theta=(\mu,Q)$}{}} \label{section:remaining_upd_elems}

In this section we first formulate the new prior for $\theta=(\mu,Q)$, where $Q$ is restricted to be
a band matrix. Thereafter we formulate a computationally efficient procedure to generate a sample from the resulting
$f_{\theta\vert z^{(m)},y}(\theta\vert z^{(m)},y)$. We start by formulating the prior.

\subsection{Prior for \texorpdfstring{$\theta=(\mu,Q)$}{}}\label{section:new_prior}

To obtain a band structure for the precision matrix $Q$ we restrict the assumed Gaussian distributions in
\eqref{prior_model} and \eqref{eq:prior} to be a Gaussian partially ordered Markov model, a Gaussian POMM
\citep{CressieDavidson1998}. To be able to give a mathematically precise definition of the Gaussian POMM we first introduce some notation.

We let $x^k$ denote the $k$'th element of $x$, so $x = (x^1,\ldots,x^{n_x})$. To each element $x^k$ of $x$
we associate a sequential neighbourhood $\Lambda_k\subseteq\{ 1,\ldots,k-1\}$, and use the notation
introduced in Section \ref{section:norm_dist_prec_matr} to denote the elements in $x$ associated to
$\Lambda_k$ by $x^{\Lambda_k}$. The number of elements in $\Lambda_k$ we denote by $\vert \Lambda_k\vert $. Moreover, we
let $\Lambda_k(1)$ denote the smallest element in the set $\Lambda_k$, we let $\Lambda_k(2)$ be the second smallest
element in $\Lambda_k$, and so on until $\Lambda_k(\vert \Lambda_k\vert )$, which is the largest element in $\Lambda_k$.
We let the distribution of $x$ be specified by the two parameter vectors $\eta=(\eta^1,\ldots,\eta^{n_x})$ and
$\phi = (\phi^1,\ldots,\phi^{n_x})$, where for each $k=1,\ldots,n_x$ we have that $\eta^k\in\mathbb{R}^{\vert \Lambda_k\vert +1}$
and $\phi^k>0$ is a scalar. With this notation the Gaussian POMM is specified as 
\begin{equation} \label{new_likelihood_joint_dist}
    f_{x\vert \eta, \phi}(x\vert \eta, \phi) = \prod_{k=1}^{n_x} f_{x^k\vert \eta^k, \phi^k, x^{\Lambda_k}}(x^k\vert \eta^k, \phi^k, x^{\Lambda_k}),
\end{equation}
where $x^k\vert \eta^k, \phi^k, x^{\Lambda_k}$ is Gaussian with mean
\begin{align}\nonumber
  \mbox{E}[x^k\vert \eta^k, \phi^k, x^{\Lambda_k}] &= \eta^{k,1}+x^{\Lambda_k(1)}\eta^{k,2}+\dots + x^{\Lambda_k(\vert \Lambda_k\vert )}\eta^{k,\vert \Lambda_k\vert +1}\\
  &= \left[\begin{array}{cc}1 & (x^{\Lambda_k})^T\end{array}\right] \eta^k
\end{align}
and variance $\mbox{Var}[x^k\vert \eta^k, \phi^k, x^{\Lambda_k}] = \phi^k$.

It should be noted that $\theta=(\mu,Q)$ in ${\mathcal N}(x;\mu,Q)$ is uniquely specified by $\eta$ and $\phi$. {If the number of sequential neighbours is small, the resulting precision matrix $Q$ becomes sparse, see  Appendices \ref{section:derivation_prec_mx} and \ref{section:seq_neigh_sparsity_Q_app}
for a detailed discussion.} We can therefore specify a
prior for $\theta$ by specifying a prior for $\eta$ and $\phi$, which we choose as conjugate to the Gaussian POMM just defined. More specifically, we first assume the elements in $\phi$ to be independent,
and each element $\phi^k$ to be inverse Gamma distributed
with parameters $\alpha^k$ and $\beta^k$. Next, we assume the elements of $\eta$ to be conditionally independent
and Gaussian distributed given $\phi$,
\begin{equation}
  \eta^k\vert \phi \sim {\mathcal N}(\eta^k;\zeta^k, (\phi^k\Sigma_{\eta^k})^{-1}),
\end{equation}
where $\zeta^k\in\mathbb{R}^{\vert \Lambda_k\vert +1}$ and $\Sigma_{\eta^k}\in\mathbb{R}^{(\vert \Lambda_k\vert +1)\times (\Lambda_k\vert +1)}$
are hyperparameters that have to be set.

\subsection{Sampling from \texorpdfstring{$f_{\theta\vert z^{(m)},y}(\theta\vert z^{(m)},y)$}{}} \label{section:par_sim}
To sample from $f_{\theta\vert z^{(m)},y}(\theta\vert z^{(m)},y)$ we adopt the same general strategy as proposed in
\citet{LoeTjelmeland2021}. We include the underlying state vector at time $t$, $x$, as an auxiliary
variable and simulate $(\theta,x)$ from 
\begin{align}\nonumber
  f_{\theta,x\vert z^{(m)},y}(\theta,x\vert z^{(m)},y) \propto f_{\theta}(\theta)f_{x\vert \theta}(x\vert \theta) f_{y\vert x}(y\vert x)\prod_{i\neq m}
  f_{x^{(i)}\vert \theta}(x^{(i)}\vert \theta)& \\ \label{eq:joint}
  = f_{\theta}(\theta)\ {\mathcal N}(x;\mu,Q)\ {\mathcal N}(y;Hx,R)\ \prod_{i\neq m}{\mathcal N}(x^{(i)};\mu,Q).&
\end{align}
By thereafter ignoring the simulated $x$ we have a sample of $\theta$ from the desired distribution. To simulate from the joint
distribution $f_{\theta,x\vert z^{(m)},y}(\theta,x\vert z^{(m)},y)$ we adopt a two block Gibbs sampler, alternating between
drawing $x$ and $\theta$ from the full conditionals $f_{x\vert \theta,z^{(m)},y}(x\vert \theta,z^{(m)},y)$ and
$f_{\theta\vert x,z^{(m)},y}(\theta\vert x,z^{(m)},y)$, respectively. We initialise the Gibbs sampler by setting $x=\frac{1}{{\mathcal M}-1} \sum_{i \neq m} x^{(i)}$. This initial value should be centrally located in $f_{x\vert z^{(m)},y}(x\vert z^{(m)},y)$, and since the Gibbs sampler we are using only consists of two blocks we should expect it to converge very fast. So just a few iterations should suffice.

From \eqref{eq:joint} we get the full conditional for $x$
\begin{equation}
  f_{x\vert \theta,z^{(m)},y}(x\vert \theta,z^{(m)},y) = f_{x\vert \theta,y}(x\vert \theta,y) \propto {\mathcal N}(x;\mu,Q)\ {\mathcal N}(y;Hx,R).
\end{equation}
It is straightforward to show that this is a Gaussian distribution with mean $\widetilde{\mu}$ and covariance matrix
$\widetilde{Q}$ given by
\begin{align}
  \widetilde{\mu} &= \mu + (Q + H^TRH)^{-1}H^T R (y-H\mu), \label{mu_tilde}\\
  \widetilde{Q} &= Q + H^TRH.\label{Q_tilde}
\end{align}
{As discussed in Appendices \ref{section:derivation_prec_mx} and \ref{section:seq_neigh_sparsity_Q_app}, the $Q$ becomes sparse whenever the number of sequential neighbours is small.} If $x$ represents values in a two dimensional lattice and the
sequential neighbourhoods $\Lambda_k$ are chosen as translations of each other, as we use in the simulation example in
Section \ref{section:sim_examples}, the $Q$ is a band matrix, {again see the discussion in Appendix
\ref{section:seq_neigh_sparsity_Q_app}}. Assuming also $R$
and $H$ to be band matrices, the product $H^TRH$ can be efficiently computed and is also a band matrix. The
$\widetilde{Q}$ is thereby also a band matrix, so the Cholesky decomposition of it can be computed efficiently as
discussed in Section \ref{section:num_prop_sparse_matrix}. The band structures of $H$ and $R$ can be used to
evaluate the right hand side of \eqref{mu_tilde} efficiently, and in addition computational efficiency can be gained by computing the product $H^TR(y-H\mu)$ in the right order. In general, multiplying two matrices $C \in \mathbb{R}^{u\times v}$ and $D \in \mathbb{R}^{v\times w}$ has computational complexity $\mathcal{O}(uvw)$. Hence, we should first compute $R(y-H\mu)$ before calculating $H^T R(y-H\mu)$. Having $\widetilde{Q}$ and the Cholesky decomposition of $\widetilde{Q}$ we can both get $\widetilde{\mu}$
and generate a sample from the Gaussian distribution efficiently as discussed in
Section \ref{section:num_prop_sparse_matrix}.

From \eqref{eq:joint} we also get the full conditional for $\theta$,
\begin{align}\nonumber
  f_{\theta\vert x,z^{(m)},y}(\theta\vert x,z^{(m)},y) &= f_{\theta\vert x,z^{(m)}}(\theta\vert x,z^{(m)}) \\
  &\propto
  f_{\theta}(\theta)\ {\mathcal N}(x;\mu,Q)\ \prod_{i\neq m}{\mathcal N}(x^{(i)};\mu,Q).
\end{align}
To simulate from this full conditional we exploit that $\theta=(\mu,Q)$ is uniquely given by the parameters $\phi$ and
$\eta$ of the Gaussian POMM prior, which means that we can first simulate values for $\phi$ and $\eta$ from
$f_{\phi,\eta\vert x,z^{(m)}}(\phi,\eta\vert x,z^{(m)})$ and thereafter use the generated $\phi$ and $\eta$ to compute the
corresponding $\mu$ and $Q$. In Appendix \ref{section:posterior_derivation_appendix} we study in detail
the resulting $f_{\phi,\eta\vert x,z^{(m)}}(\phi,\eta\vert x,z^{(m)})$ and show that it has the same form as the
corresponding prior $f_{\phi,\eta}(\phi,\eta)$, but with updated parameters. More specifically, the elements of
$\phi$ are conditionally independent given $x$ and $z^{(m)}$, with 
\begin{equation}\label{phi_marginal_posterior}
    \phi^k\vert z^{(m)},x \sim \text{InvGam}(\widetilde{\alpha}^k, \widetilde{\beta}^{(m),k}),
\end{equation}
where
\begin{align}
    \widetilde{\alpha}^k &= \alpha^k+\frac{{\mathcal M}}{2},\label{alpha_phi_posterior}\\
    \widetilde{\beta}^{(m),k} &= \left(\frac{1}{\beta^k}+\frac{1}{2}(\gamma^{(m),k}-(\rho^{(m),k})^T(\Theta^{(m),k})^{-1}\rho^{(m),k})\right)^{-1},\label{beta_phi_posterior}
\end{align}
\begin{align}
    \gamma^{(m),k} &= (\zeta^k)^T \Sigma_{\eta^k}^{-1}\zeta^k+(\chi^{(m),k})^T \cdot \chi^{(m),k}, \label{gamma_k}\\
    \rho^{(m),k} &= \Sigma_{\eta^k}^{-1}\zeta^k + (1,(\chi^{(m),\Lambda_k})^T)^T \cdot  (\chi^{(m),k})^T, \label{rho_k} \\
    \Theta^{(m),k} &= \Sigma_{\eta^k}^{-1}+(1,(\chi^{(m),\Lambda_k})^T)^T \cdot  (1,(\chi^{(m),\Lambda_k})^T)\label{Theta_k}
\end{align}
and
\begin{align}
    \chi^{(m),k} &= (x^{(1),k}, \dots , x^{(m-1),k}, x^{(m+1),k}, \dots , x^{({\mathcal M}),k}, x^k),\\
    \chi^{(m),\Lambda_k} &= (x^{(1),\Lambda_k}, \dots , x^{(m-1),\Lambda_k}, x^{(m+1),\Lambda_k}, \dots , x^{({\mathcal M}),\Lambda_k}, x^{\Lambda_k}).
\end{align}
The distribution for $\eta$ given $\phi$, $x$ and $z^{(m)}$ becomes
\begin{equation} \label{eta_joint_posterior}
    f_{\eta\vert \phi, z^{(m)},x}(\eta\vert \phi, z^{(m)},x) = \prod_{k=1}^{n_x} f_{\eta^k \vert \phi^k, z^{(m)}, x}(\eta^k \vert \phi^k, z^{(m)}, x)
\end{equation}
where
\begin{equation} \label{eta_marginal_posterior}
    \eta^k \vert \phi^k, z^{(m)}, x \sim {\mathcal N}(\eta^k; (\Theta^{i,k})^{-1}\rho^{i,k}, (\phi^k)^{-1} \Theta^{i,k}).
\end{equation}
In particular we see that it is easy to sample from $f_{\phi,\eta\vert x,z^{(m)}}(\phi,\eta\vert x,z^{(m)})$ by first sampling
the elements of $\phi$ independently according to \eqref{phi_marginal_posterior} and thereafter generate
the elements of $\eta$ according to \eqref{eta_marginal_posterior}. Having samples of $\phi$ and $\eta$ we can thereafter
compute the corresponding $\mu$ and $Q$ as detailed in Appendix \ref{section:derivation_prec_mx}.

\section{Block update} \label{section:block_upd}
Section \ref{section:bayes_upd_ensemble} presents a set of updating procedures that allows us to update a prior realisation into a posterior realisation, where the posterior realisation takes the observation in the current time step into account. In Section \ref{section:opt_sol}, we found the optimal filter according to our chosen criterion. However, as also discussed in Section \ref{section:res_upd_proc}, parts of this procedure is computationally demanding. In the following we introduce the approximate, but computationally more efficient procedure for Steps \ref{alg:svd1} to {\ref{alg:update_x^m}} in Algorithm \ref{alg:optimal_update} that we mentioned in Section \ref{section:res_upd_proc}. So the situation considered {is that we want to update a prior realisation $x^{(m)}$ and} that we already have generated a mean vector $\mu$ and a sparse precision matrix $Q$. The goal is
now to define an approximation to the { vector $\widetilde{x}^{(m)}$} given in Step {7} in Algorithm \ref{alg:optimal_update}.
The strategy is then to use the approximation to {$\widetilde{x}^{(m)}$}, which we denote by {$\widetilde{x}^{(m),\mbox{\scriptsize Block}}$},
instead of {$\widetilde{x}^{(m)}$}. In the following we assume the matrices $H$ and $R$ to be sparse.

{The complete block updating procedure we propose is summarised in Algorithm \ref{alg:block_update}. In the following we motivate and
  explain the various steps, and define the notation used.}
\begin{algorithm}[tbp]
    {
      \begin{enumerate}
      \item[] \hspace*{-0.45cm}Given $x^{(1)},\ldots,x^{({\mathcal M})}$, $y$, $H$, $R$ and $C_b$, $D_b$, $E_b$ for $b=1,\ldots,{\cal B}$.
      \item[] \hspace*{-0.45cm}For $m=1,\ldots,{\mathcal M}$:
      \item \hspace*{0.25cm}Sample $\theta = (\mu,Q)$ from $f_{\theta\vert z^{(m)},y}(\theta\vert z^{(m)},y)$.
      \item \hspace*{0.25cm}For $b=1,\ldots,{\cal B}$:
        \begin{enumerate}
        \item[\hspace*{0.45cm}{\bf a}.]\hspace*{0.25cm} Form the set $J_b$ in \eqref{eq:Jb}.
        \item[\hspace*{0.45cm}{\bf b}.]\hspace*{0.25cm} Form $g(x^{E_b}, y^{J_b}\vert \theta)$ in \eqref{eq:g1}.
        \item[\hspace*{0.45cm}{\bf c}.]\hspace*{0.25cm} Form $g(x^{D_b}, y^{J_b}\vert\theta)$ by marginalising $g(x^{E_b},y^{J_b}\vert\theta)$ over $x^{E_b\setminus D_b}$.
        \item[\hspace*{0.45cm}{\bf d}.]\hspace*{0.25cm} Form $g(x^{D_b}\vert \theta)$ and $g(y^{J_b}\vert\theta,x^{D_B})$ from $g(x^{D_b},y^{J_b}\vert\theta)$, and
          identify $\widetilde{\mu}^b$, 
          \hspace*{0.35cm}$\widetilde{Q}^b$, $\widetilde{a}^b$, $\widetilde{H}^b$ and $\widetilde{R}^b$.
        \item[\hspace*{0.45cm}{\bf e}.]\hspace*{0.25cm} Compute $\widetilde{x}^{(m),D_b}$ by Steps 2 to 7 in Algorithm \ref{alg:optimal_update}
          when $x^{(m)}$, $\widetilde{x}^{(m)}$, \newline \hspace*{0.25cm} $y$, $\mu$, $Q$, $H$ and $R$ are replaced with
          $x^{(m),D_b}$, $\widetilde{x}^{(m),D_b}$, $y^{J_b}-\widetilde{a}^b$, $\widetilde{\mu}^b$, \newline \hspace*{0.35cm}$\widetilde{Q}^b$, $\widetilde{H}^b$ and $\widetilde{R}^b$,
          respectively.
        \end{enumerate}
      \item \hspace*{0.45cm}Define $\widetilde{x}^{(m),\mbox{\scriptsize Block}}$ by setting $\widetilde{x}^{(m),\mbox{\scriptsize Block}}_\ell=\widetilde{x}^{(m),D_b}_{\ell_b}$ for
        $\ell\in C_b, \newline \hspace*{0.45cm} b=1,\ldots,{\cal B}$.
      \end{enumerate}
      \caption{Summary of the block updating procedure for computing the posterior ensemble
        $\widetilde{x}^{(1),\mbox{\scriptsize Block}},\ldots,\widetilde{x}^{({\mathcal M}),\mbox{\scriptsize Block}}$ from the prior ensemble $x^{(1)},\ldots,x^{({\mathcal M})}$.}
      \label{alg:block_update}
    }
    \end{algorithm}

Our construction of {$\widetilde{x}^{(m),\mbox{\scriptsize Block}}$} is general, but to motivate our construction we consider a situation where the elements of the state vector are related to nodes in a two-dimensional (2D) lattice of nodes, with $r$ rows and $s$ columns say, and where the correlation between two elements of the state vector decay with distance between the corresponding nodes. We assume the nodes in the lattice to be numbered in the lexicographical order, so node $(k, \ell)$ in the lattice is related to the value of element number $(k-1)\cdot s+\ell$ of the state vector. We use ${\mathcal S}=\{(k,\ell): k=1, \dots , r, l=1, \dots , s\}$ to denote the set of all indices of the elements in the state vector, where we for the 2D lattice example have $n_x=rs$. 

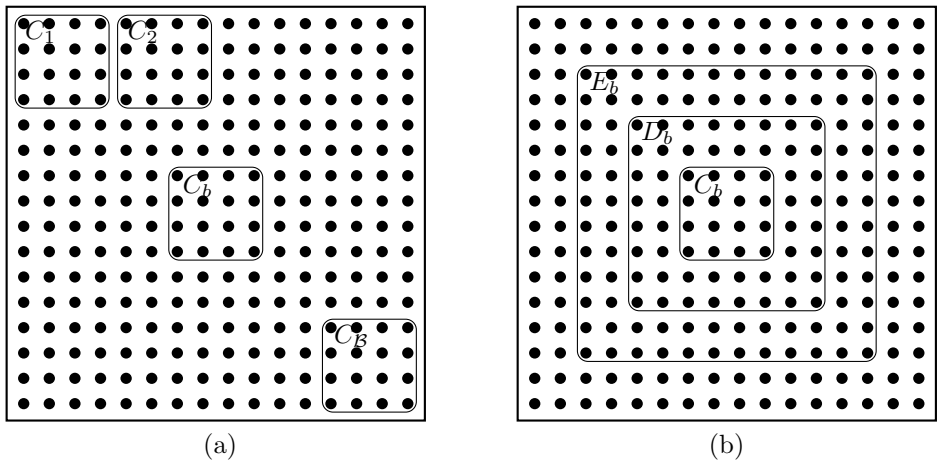
\begin{figure}
  \begin{center}
    \begin{tabular}{ccc}
      \begin{tikzpicture}[scale=0.45]
    
    \draw[black, thick] (0,0) rectangle (12.25,12.25);
   
    \draw[rounded corners] (0.25, 9.25) rectangle (3, 12) {};
    \draw[rounded corners] (3.25, 9.25) rectangle (6, 12) {};
    \draw[rounded corners] (4.75, 4.75) rectangle (7.5, 7.5) {};
    \draw[rounded corners] (9.25, 0.25) rectangle (12, 3) {};
    
    \draw[black, thick] (1,11.5) node {$C_1$};
    \draw[black, thick] (4,11.5) node {$C_2$};
    \draw[black, thick] (5.6,6.95) node {$C_b$};
    \draw[black, thick] (10.10,2.5) node {$C_{\mathcal B}$};
    
    \foreach \x in {0.5,1.25,..., 11.75}{
        \foreach \y in {0.5,1.25,..., 11.75}{
            \node at (\x,\y)[circle, fill, inner sep = 1.5pt]{};
        }
    }
\end{tikzpicture}
      & ~~
      &
      \begin{tikzpicture}[scale=0.45]
    
    \draw[black, thick] (0,0) rectangle (12.25,12.25);
   
    \foreach \x in {0.5,1.25,...,11.75}{
        \foreach \y in {0.5,1.25,..., 11.75}{
            \node at (\x,\y)[circle, fill, inner sep = 1.5pt]{};
        }
    }
    
    \draw[rounded corners] (4.75, 4.75) rectangle (7.5, 7.5) {};
    \draw[rounded corners] (3.25, 3.25) rectangle (9, 9) {};
    \draw[rounded corners] (1.75, 1.75) rectangle (10.5, 10.5) {};
    
    \draw[black, thick] (5.6,6.95) node {$C_b$};
    \draw[black, thick] (4.1,8.5) node {$D_b$};
    \draw[black, thick] (2.55,10) node {$E_b$};

\end{tikzpicture}\\
      (a) & & (b) \\
    \end{tabular}
    \end{center}
    \caption{Example of how the sets $C_b$, $D_b$ and $E_b$ can be chosen when the elements of $x$ are related to the nodes of a 2D lattice. The dots denote nodes. In (a) the nodes are partitioned in $\mathcal{B}$ different blocks, while (b) shows an example of how $D_b$ and $E_b$ might look like}
    \label{fig:block_update_setup}
\end{figure}

The first step in the construction of {$\widetilde{x}^{(m),\mbox{\scriptsize Block}}$} is to define a partition $C_1,\ldots, C_{\mathcal B}$
of ${\mathcal S}$. The sets $C_b, b=1,\ldots, {\mathcal B}$ should be chosen so that { elements
of the state vector corresponding to elements in the same set $C_b$ are highly correlated.} In the 2D lattice
example the most natural choice would be to let each $C_b$ be a block of consecutive nodes in the lattice,
for example a block of $r_C\times s_C$ nodes. Such a choice of the partition is visualised in
Figure \ref{fig:block_update_setup}(a). 
One should note that a similar partition of ${\mathcal S}$ is
also used in the domain localisation version of EnKF \citep{domain_localization_1}. However, the
motivation for the partition in domain localisation is mainly to eliminate or at least reduce
the effect of spurious correlations \citep{domain_localization_2}, whereas in our setup the
motivation is merely to reduce the computational complexity of the updating procedure. In particular,
if the dimension of the state vector, $n_x$, is sufficiently small we recommend not to use the block
updating procedure at all, one should then {rather} use the procedure summarised in
Section \ref{section:remaining_upd_elems}.

{
  When forming $\widetilde{x}^{(m)}$ from $x^{(m)}$ in Step 7 of Algorithm \ref{alg:optimal_update}
  one would intuitively expect that element number $k$ of $x^{(m)}$ has a negligible effect on element $\ell$ of
  $\widetilde{x}^{(m)}$ whenever the correlation between elements $k$ and $\ell$ of the state vector is sufficiently weak.}
In numerical experiments where the dimension of the state vector is sufficiently small so that we can use the
procedure in Section \ref{section:res_upd_proc}, this intuition has already been confirmed.
{This motivates our first approximation, which is to not allow an element of $x^{(m)}$ to influence
  an element of $\widetilde{x}^{(m),\mbox{\scriptsize Block}}$ when the two elements of the state
  vector have a very low correlation.}
More formally, for each $b=1, \ldots,{\mathcal B}$ we define a set of nodes $D_b$ so that $C_b\subseteq D_b$,
{where $D_b$ should be chosen so that in the state vector, elements $j\in C_b$ are only weakly correlated with elements $i\in {\cal S}\setminus D_b$. In the approximation we
  only allow an element $\ell\in C_b$ of $\widetilde{x}^{(m),\mbox{\scriptsize Block}}$ to be a function of 
  element $k$ of $x^{(m)}$ if $k\in D_b$.}
In the 2D lattice example it is natural to define $D_b$ by expanding the $C_b$ block of nodes with $u$, say, nodes in each direction, see the illustration in Figure \ref{fig:block_update_setup}(b) where $u=2$ is used.

To decide {how $\widetilde{x}^{(m),\mbox{\scriptsize Block}}_\ell,\ell\in C_b$ is given from $x^{(m)}_k,k\in D_b$},
the most natural procedure would be the following.
Start with the assumed joint Gaussian distribution for the state vector $x$ and the observation vector $y$,
{ \begin{equation}\label{eq:jointExact}
  f_{x,y\vert\theta}(x,y\vert\theta) = {\cal N}(x;\mu,Q) \cdot {\cal N}(y;Hx,R).
\end{equation}}
Then { we could have marginalised} out all elements of the state vector that are not in $D_b$,
{
  \begin{equation}
    \label{eq:marg}
    f_{x^{D_b},y\vert\theta}(x^{D_b},y\vert\theta) = \int f_{x,y\vert\theta}(x,y\vert\theta) dx^{-D_b},
  \end{equation}
where we use the notation introduced in Section \ref{section:norm_dist_prec_matr} and let $x^{D_b}$ and
$x^{-D_b}$ denote vectors of the elements of $x$ related to the sets $D_b$ and ${\cal S}\setminus D_b$,
respectively.}
From this, one could find the marginal distribution for the elements of the state vector $x$ related
to the block $D_b$, { i.e. $f_{x^{D_b}\vert\theta}(x^{D_b}\vert\theta)$,} and the conditional
distribution for the observation vector $y$ given the elements of the state vector related
to $D_b${, $f_{y\vert x^{D_b},\theta}(y\vert x^{x_b},\theta)$}.
{As $f_{x^{D_b}\vert\theta}(x^{D_b}\vert\theta)$ becomes a Gaussian distribution,
  and $f_{y\vert x^{D_b},\theta}(y\vert x^{x_b},\theta)$ becomes a Gaussian distribution where
  the mean vector that is a linear function of $x^{D_b}$ and the precision matrix is constant,
  one could use the procedure specified in Section \ref{section:res_upd_proc}
  to find the optimal update of the sub-vector $x^{(m),D_b}$, $\widetilde{x}^{(m),D_b}$ say.
  Finally, we could for each $\ell\in C_b$
  have set $\widetilde{x}^{(m),\mbox{\scriptsize Block}}_\ell$ equal to the the value of the corresponding
  element in $\widetilde{x}^{(m),D_b}$.}
However, such a procedure is not computationally feasible when the dimension of the state vector
is large, and for two reasons. First, the marginalisation { in \eqref{eq:jointExact}}
is computationally expensive when dependence is represented by precision matrices.
Second, when the dimension of the state vector is large the dimension of the observation vector
$y$ is typically also large, which makes { such a} marginalisation process even more
computationally expensive. We therefore do one more approximation before following the procedure
described {above}. Instead of starting out with { \eqref{eq:jointExact}},
we start with { a} corresponding conditional distribution{,}
where we condition on the value of elements of the state
{vector $x$} and observation {vector $y$}
that are only weakly correlated with the elements in {$x^{D_b}$. The motivation for
such a conditioning is twofold. First, to condition on elements that are only weakly correlated
with the elements in $x^{D_b}$ should not change the distribution of $x^{D_b}$ much. Second,
when representing dependence by precision matrices the computation of 
conditional distributions is computationally very efficient.} In the following, we define this last
approximation more formally { and discuss related computational aspects.}

\begin{sloppypar}
  For each $b=1, \dots {\mathcal B}$, we define a set of nodes $E_b$ such that $D_b \subseteq E_b$.
  The set $E_b$ should be chosen so that the nodes in $D_b$ are weakly correlated with the nodes that
  are not in $E_b$. For the 2D lattice example it is reasonable to define $E_b$ by expanding the $D_b$
  block with, say, $v$ nodes in each direction. Figure \ref{fig:block_update_setup}(b) displays an
  example of $E_b$ where $v=2$. Moreover, we {let} a set $J_b$ contain the indices of the
  elements in $y$ that { by the likelihood} are linked to
  {one or more elements in} $x^{E_b}$, {i.e.
  \begin{equation}\label{eq:Jb}
    J_b = \{k \in {\mathcal S}: H^{k,\ell}\neq 0 \mbox{~for at least one $(k,\ell)$}, k\in \{1,\ldots,n_y\},\ell\in E_b\}.
  \end{equation}}
{Starting from \eqref{eq:jointExact} we now find the corresponding
    conditional distribution when conditioning on $x^k$ being equal to its mean value $\mu^k$ for all
    $k \not\in E_b$, and conditioning on $y_k$ being equal to its mean value $(Hx)^k$ for all $k\not\in J_b$.
    The resulting conditional distribution we denote by} 
  \begin{equation}\label{eq:g1}
    g(x^{E_b}, y^{J_b}\vert \theta) = f(x^{E_b}, y^{J_b}\vert \theta, x^{-E_b}=\mu^{-E_b}, y^{-J_b}=(Hx)^{-J_b}).
  \end{equation}
  {This (conditional) distribution is also multivariate Gaussian, and the mean vector and the
    precision matrix can be found by using the expressions} in Section \ref{section:norm_dist_prec_matr}.
  {One should note that the conditional precision matrix is immediately given, simply by
    clipping out the parts of the unconditional precision matrix that belong to $x^{E_b}$ and $y^{J_b}$.
    The formula for the conditional mean includes a matrix inversion, where the size of the matrix
    that needs to be inverted equals the sum of the dimensions of the vectors $x^{E_b}$ and $y^{J_b}$.
    The dimension of the matrix is thereby not problematically large. Moreover, one should note that in
    practice the conditional mean is computed using the techniques discussed in Section
    \ref{section:num_prop_sparse_matrix}, thereby avoiding the matrix inversion.}
  From $g(x^{E_b}, y^{J_b}\vert \theta)$ we marginalise over $x^{E_b\setminus D_b}$ to get $g(x^{D_b},y^{J_b}\vert \theta)$,
  which is also Gaussian and where the mean vector and the precision matrix can be found as discussed
  in Section \ref{section:norm_dist_prec_matr}. {The $g(x^{D_b},y^{J_b}\vert \theta)$ should be 
    thought of as a substitute for the $f(x^{D_b},y^{J_b}\vert\theta)$ defined in (\ref{eq:marg}). So following what we
    ideally would have liked to do with $f(x^{D_b},y^{J_b}\vert\theta)$ we use $g(x^{D_b},y^{J_b}\vert \theta)$ to form
    the marginal distribution for $x^{D_b}$, $g(x^{D_b}\vert\theta)$, and the conditional distribution for
    $y^{J_b}$ given $x^{D_b}$, $g(y^{J_b}\vert\theta,x^{D_b})$. These two distributions are both Gaussian and
    can} be found using the expressions in Section \ref{section:norm_dist_prec_matr}.
  {The $g(x^{D_b}\vert\theta)$ should be thought of as a prior for $x^{D_b}$, and the $g(y^{J_b}\vert\theta,x^{D_b})$
    represents the likelihood for $y^{J_b}$ given $x^{D_b}$. In the following we let $\widetilde{\mu}^b$ and
    $\widetilde{Q}^b$ denote the resulting mean vector and precision matrix in the prior $g(x^{D_b}\vert\theta)$,
    respectively. From (\ref{cond_dist_norm_prec}) we get that the Gaussian 
    likelihood $g(y^{J_b}\vert\theta,x^{D_b})$ has a conditional mean in the form
    $\mbox{E}[y^{J_b}\vert x^{D_b},\theta]=\widetilde{a}^b+\widetilde{H}^bx^{D_b}$,
    where $\widetilde{a}^b$ is a column vector of size $\vert J_b\vert$ and $\widetilde{H}^b$ is a $\vert J_b\vert\times \vert D_b\vert$ matrix.
    The precision matrix of the likelihood $g(y^{J_b}\vert\theta,x^{D_b})$ we in the following denote by
    $\widetilde{R}^b$. Noting that observing $y^{J_b}$ is equivalent to observing
    $y^{J_b}-\widetilde{a}^b$, we thereby have a Bayesian model corresponding to what we discussed in
    Section \ref{section:bayesian_model}, except that now $x$ is replaced with $x^{D_b}$, the 
    ensemble element $x^{(m)}$ is similarly replaced with the sub-vector $x^{(m),D_b}$,
    and $y$, $\mu$, $Q$, $H$ and $R$ are replaced with $y^{J_b}-\widetilde{a}^b$, $\widetilde{\mu}^b$,
    $\widetilde{Q}^b$, $\widetilde{H}^b$ and $\widetilde{R}^b$,
    respectively. To update $x^{(m),D_b}$ we can thereby use Steps 2 to 7 in Algorithm \ref{alg:optimal_update}
    to find the optimal update of $x^{(m),D_b}$, $\widetilde{x}^{(m),D_b}$ say. Finally, for each $\ell\in C_b$,
    we set $\widetilde{x}^{(m),\mbox{\scriptsize Block}}_\ell = \widetilde{x}^{(m),D_b}_{\ell_b}$, where $\ell_b$ is the index of
    the element in $\widetilde{x}^{(m),D_b}$ that corresponds to element $\ell$ in $\widetilde{x}^{(m),\mbox{\scriptsize Block}}$.}


\end{sloppypar}

\section{Simulation example} \label{section:sim_examples} 
{In the following we present a numerical experiment with the model-based EnKF
  using the proposed block update approximations.
  The purpose of the experiment is both to study the 
  computational speedup we obtain by using the block updating, and to evaluate empirically
  the quality of approximation introduced by adopting the block updating procedure.
  To do this we do a simulation study where we run the model-based EnKF with block updates
  and the exact model-based EnKF on the same observed values, and compare the run time
  and filter outputs. As our focus is on the effect of the
  approximation introduced by using the block update, we use the Gaussian POMM prior distribution
  introduced in Section 4.1 for both filter variants and we also use the exact same observed data
  in both cases.} {We first define and study the results for a linear example in Section
  \ref{section:lin_example} and thereafter discuss 
  a non-linear examples in Section \ref{section:non_lin_example}.}

\subsection{{Linear example}}\label{section:lin_example}

{In the simulation experiment} we first generate a series of reference states $\{x_t\}_{t=1}^T$ and corresponding observed values $\{y_t\}_{t=1}^T$. The series of reference states we consider as the unknown
true values of the latent $x_t$ process and we use the corresponding
generated observed values in the two versions of EnKF.
We then compare the computational demands and the output of the two filtering
procedures. {Lastly, we compare the output of the block update procedure to the Kalman filter solution.}

\subsubsection{Reference time series and observations}\label{section:lin:ref_sol_and_obs}
\begin{figure}
  \begin{center}
    \begin{tabular}{ccc}
      \includegraphics[width=5.1cm, height=4.0cm]{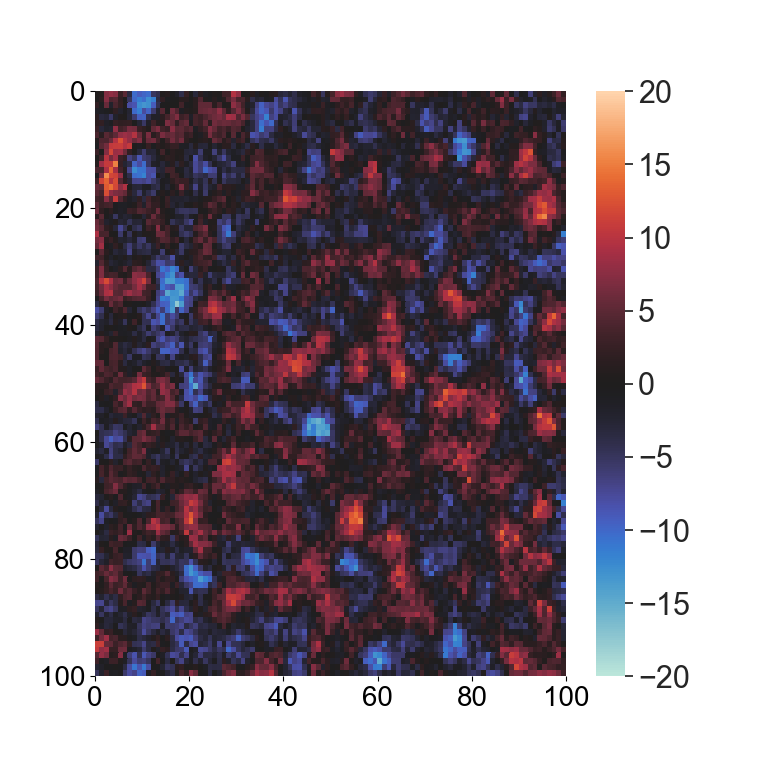}
      & ~~
      &
      \includegraphics[width=5.1cm, height=4.0cm]{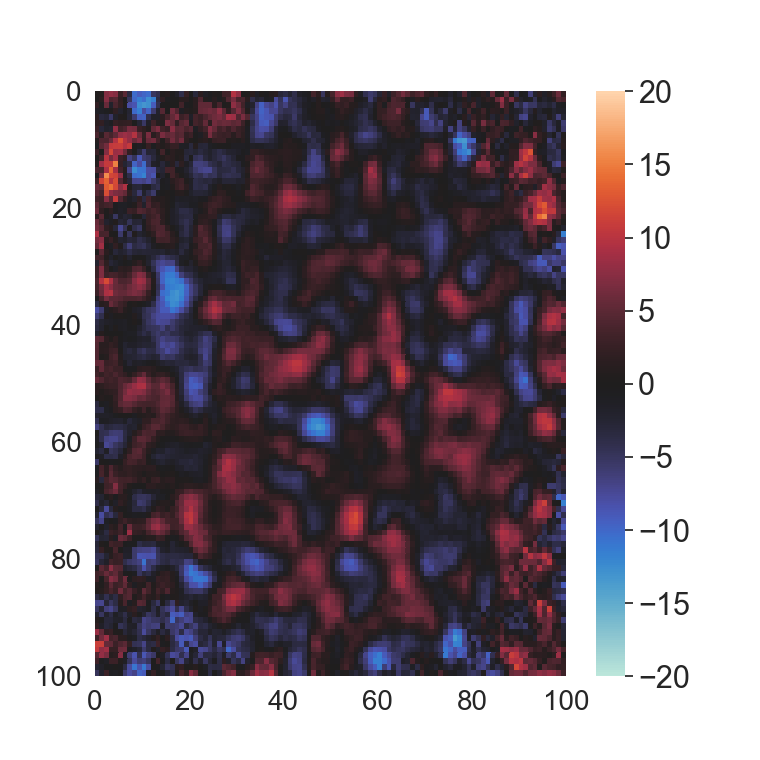} \\[-0.2cm]
      (a) & & (b) \\[-0.1cm]
      \includegraphics[width=5.1cm, height=4.0cm]{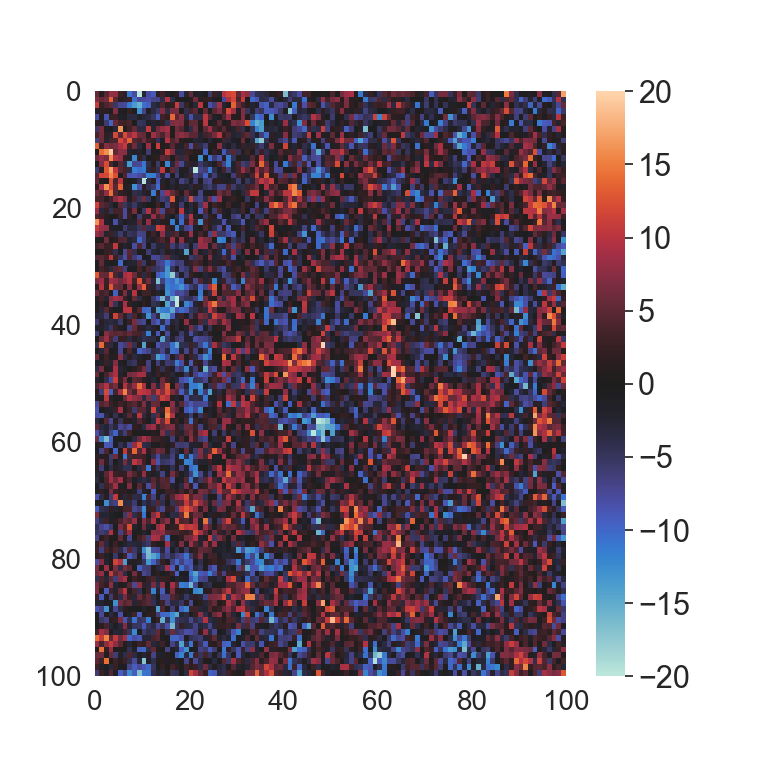}
      & &
      \includegraphics[width=5.1cm, height=4.0cm]{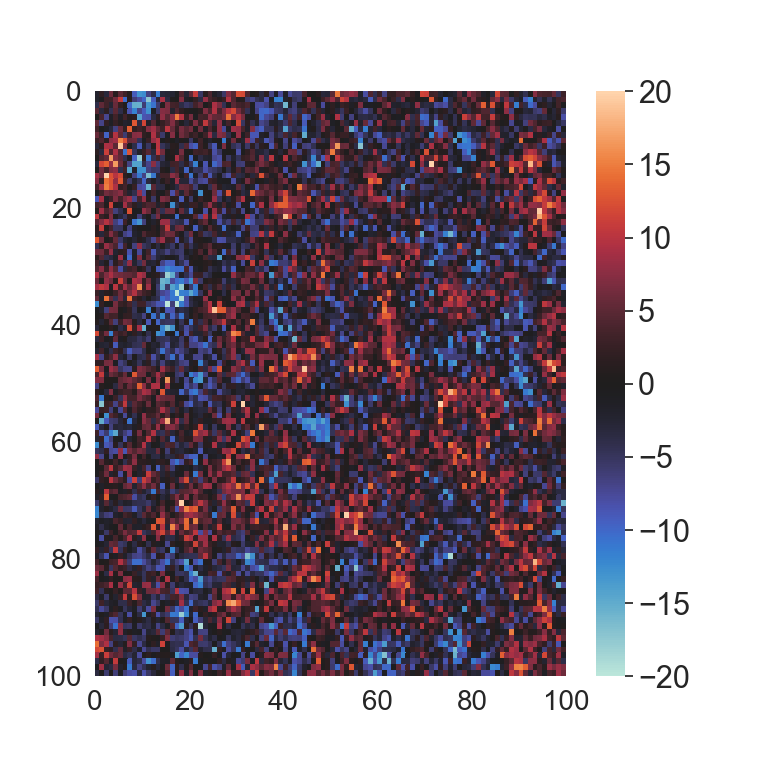} \\[-0.2cm]
      (c) && (d)\\
    \end{tabular}
  \end{center}
  \caption{{Linear example:} Reference states at times (a) $t=1$ and (b) $t=5$, and corresponding generated observed values at
    times (c) $t=1$ and (d) $t=5$}
  \label{fig:ref_sol_and_data}
\end{figure}

To generate the series of reference states we adopt a two dimensional variant of the one dimensional
setup used in \citet{HEnKF}. So for each time step $t=1,\ldots,T$ the reference state $x_t$ represents
the values in a two dimensional $s\times s$ lattice. As described in Section \ref{section:block_upd} we
number the nodes in the lexicographical order. The reference state at time $t=1$ we generate as a moving average of a white noise field. More precisely, we first generate independent and standard normal variates
$z^{(k,\ell)}$ for each node $(k,\ell)$, and to avoid boundary effects we do this for an extended lattice. The
reference state at node $(k,\ell)$ and time $t=1$ is then defined by
\begin{equation}
x_1^{(k-1)\cdot s + \ell} = \sqrt{\frac{20}{\vert \Gamma_{3,(k,\ell)}\vert }}\sum_{(i,j)\in\Gamma_{3,(k,\ell)}} z^{(i,j)},
\end{equation}
where $\Gamma_{r,(k,\ell)}$ is the set of nodes in the extended lattice, that are less than or equal to a distance 
$r$ from node $(k,\ell)$, and $\vert \Gamma_{r,(k,\ell)}\vert $ is the  number of elements in that set.
One should note that the factor $\sqrt{20/\vert \Gamma_{3,(k,\ell)}\vert }$ gives that the variance of each
generated $x_1^{(k-1)\cdot s + \ell}$ is $20$. Figure \ref{fig:ref_sol_and_data}(a)
shows a generated $x_1$ with $s=100$, which is used in the numerical experiments discussed in
Sections \ref{section:lin:block_vs_opt} and \ref{section:block_upd_vs_KF}.

Given the reference state at time $t=1$, corresponding reference states at later time steps are
generated sequentially. For $t=2,\ldots,T$ the reference state at time $t-1$ is used to generate the reference
state at time $t$ by performing a moving average operation on nodes that are inside an annulus defined
by circles centred at the middle of the lattice. For $t=2$ the radius of the inner circle defining the
annulus is zero, and
as $t$ increases the annulus is gradually moved further away from the centre. More precisely, to generate $x_t$ from
$x_{t-1}$ we define the annulus by the two radii
$r_1 = \max\{0, \lfloor{\left(\frac{s}{2}-1\right)\frac{1}{T-1}\left(t-\frac{5}{2}\right)\rfloor}\}$ and
$r_2= \lfloor{}\left(\frac{s}{2}-1\right)\frac{t-1}{T-1}\rfloor{}$, where $\lfloor v \rfloor$ denotes the largest
integer less than or equal to $v$. For all nodes $(k,\ell)$ inside the annulus we define
\begin{equation}
    x_t^{(k-1)\cdot s+\ell} = \frac{1}{{\vert \Gamma_{1,(k,\ell)}\vert }} \sum_{(i,j)\in \Gamma_{1,(k,\ell)}} x_{t-1}^{(i-1)\cdot s+j}.
\end{equation}
For nodes $(k,\ell)$ that are not inside the specified annulus the values are unchanged, i.e. for such nodes we have
$x_t^{(k-1)\cdot s+\ell}=x_{t-1}^{(k-1)\cdot s+\ell}$. The reference solution at time $T=5$ corresponding to the
$x_1$ given in Figure \ref{fig:ref_sol_and_data}(a) is shown in Figure \ref{fig:ref_sol_and_data}(b). By comparing
the two we can see the effect of the smoothing operations.

For each time step $t=1,\ldots,T$ we generate an observed value associated to each node in the lattice, so $n_y=n_x$.
We generate the observed values at time $t$, $y_t$, by blurring $x_t$ and adding independent Gaussian noise
in each node. More precisely, the observed value in node $(k,\ell)$ at time $t$ is generated as
\begin{equation}\label{eq:obs}
  y_t^{(k-1)\cdot s+\ell}=\frac{1}{\vert \Gamma_{\sqrt{2},(k,\ell)}\cap \mathcal{S}\vert } \sum_{(i,j)\in\Gamma_{\sqrt{2},(k,\ell)}\cap \mathcal{S}}
  x_t^{(i-1)\cdot s + j}+w^{(k-1)\cdot s+\ell},
\end{equation}
where $w^{(k-1)\cdot s+\ell}$ is a zero-mean normal variate with variance $20$. 
Figures \ref{fig:ref_sol_and_data}(c) and \ref{fig:ref_sol_and_data}(d) shows the generated observations
corresponding to the $x_1$ and $x_5$ in Figures \ref{fig:ref_sol_and_data}(a) and (b), respectively.
These observations are used in the numerical experiments discussed in Sections \ref{section:lin:block_vs_opt} and \ref{section:block_upd_vs_KF}.

\subsubsection{Assumed model and algorithmic parameters}\label{section:num_exp_assumed_model}
In the numerical experiments with EnKFs we use the assumed model defined in Section \ref{section:bayesian_model}.
For nodes sufficiently far away from the lattice borders we let the sequential neighbourhood consist
of ten nodes as shown in Figure \ref{fig:seq_neigh_num_exp}. 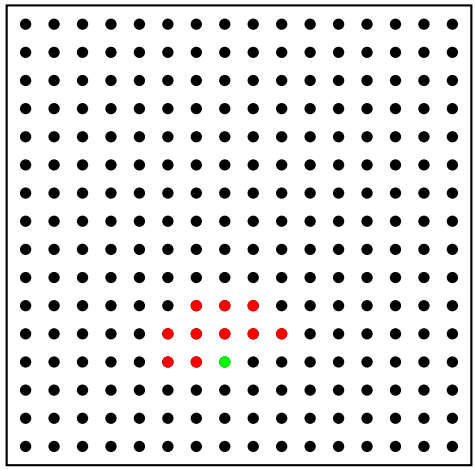
\begin{figure}
    \begin{center}
        \begin{tikzpicture}[scale=0.5]
            
            \draw[black, thick] (0,0) rectangle (12.25,12.25);

            \foreach \x in {0.5,1.25,..., 11.75}{
                \foreach \y in {0.5,1.25,..., 11.75}{
                    \node at (\x,\y)[circle, fill, inner sep = 1.5pt]{};
                }
            }
            
        \node at (5.75, 2.75)[circle, fill, inner sep = 1.5pt, green]{};
        \node at (5, 2.75)[circle, fill, inner sep = 1.5pt, red]{};
        \node at (4.25, 2.75)[circle, fill, inner sep = 1.5pt, red]{};
        \node at (4.25, 3.50)[circle, fill, inner sep = 1.5pt, red]{};
        \node at (5, 3.50)[circle, fill, inner sep = 1.5pt, red]{};
        \node at (5.75, 3.50)[circle, fill, inner sep = 1.5pt, red]{};
        \node at (6.5, 3.50)[circle, fill, inner sep = 1.5pt, red]{};
        \node at (7.25, 3.50)[circle, fill, inner sep = 1.5pt, red]{};
        \node at (5, 4.25)[circle, fill, inner sep = 1.5pt, red]{};
        \node at (5.75, 4.25)[circle, fill, inner sep = 1.5pt, red]{};
        \node at (6.5, 4.25)[circle, fill, inner sep = 1.5pt, red]{};
        \end{tikzpicture}

    \end{center}
    
    \caption{The sequential neighbourhood used in the numerical examples. The red dots denote the sequential neighbours of the green node}
    \label{fig:seq_neigh_num_exp}
\end{figure} This set of sequential neighbours should be
sufficient to be able to represent a large variety of spatial correlation structures
at the same time as it induces sparsity in the resulting precision matrix. For nodes close to the lattice borders we reduce the
number of sequential neighbours to consist of only the subset of the ten nodes shown in Figure
\ref{fig:seq_neigh_num_exp} that are within the lattice.

As priors for $\eta$ and $\phi$ we use the parametric forms specified in Section \ref{section:new_prior}.
We want these priors to be vague so that
the resulting posterior distributions are mainly influenced by the prior ensemble. We let the prior
for $\phi$ be improper by setting $\alpha^k = 0$ and $\beta^k = \infty$ for all $k$. In the prior for
$\eta_k\vert \phi_k$ we set $\zeta^k = 0$ and $\Sigma_{\eta^k}=100\cdot I_{\vert \Lambda_k\vert +1}$ for all $k$.

In preliminary simulation experiments we found that the Gibbs sampler for the sampling of $\theta$ converged
very rapidly, consistent with our discussion in Section \ref{section:par_sim}. In the numerical experiments we
therefore only used five iterations of the Gibbs sampler. When using the block update procedure we let each
$C_b$ consist of a block of $20\times 20$ nodes. To define the corresponding sets $D_b$ and $E_b$ we follow the
procedure outlined in Section \ref{section:block_upd} with $u=v=5$.

\subsubsection{Comparison of computational demands}\label{section:result_ex_3}
The main objective of the block update procedure is to provide a computationally efficient approximation
to the optimal model-based EnKF procedure defined in Section \ref{section:bayes_upd_ensemble}.
In this section we compare the computational demands of the two updating procedures as a function of the number of nodes in the lattice.

\begin{figure}
  \begin{center}
    \includegraphics[width=10.0cm,height=4cm]{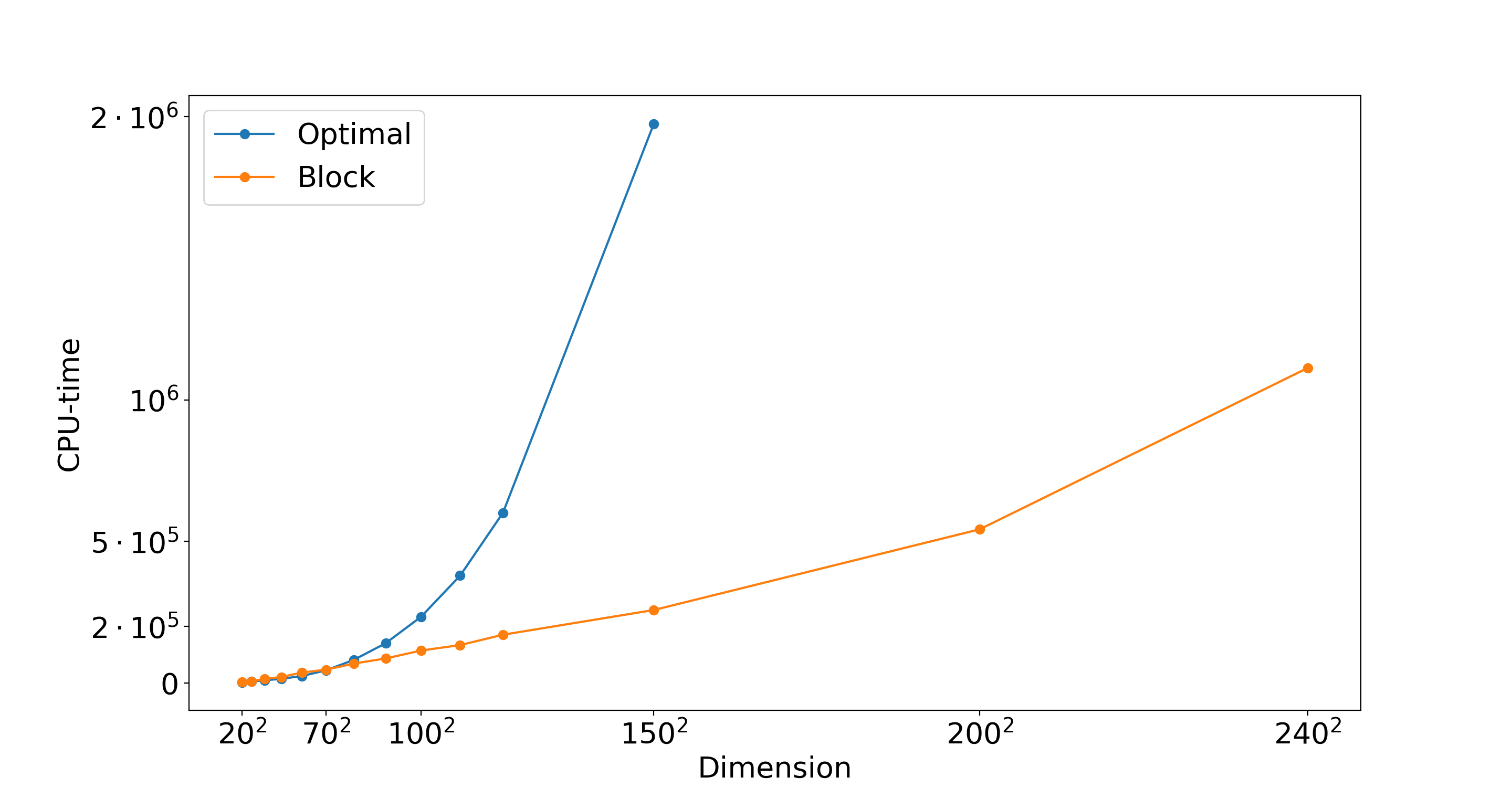}
    \caption{{Linear example:} Computing time used for running each of the
      two model-based EnKF procedures with ${\mathcal M}=25$ ensemble elements for $T=5$ time steps as a function of the number of nodes $n_x=s^2$ in the lattice.
      Computing time for the optimal model-based EnKF procedure is shown in blue, and corresponding computing time for
      the block updating procedure is shown in orange}
      \label{fig:CPU_time_comparison} 
  \end{center}
\end{figure}

For an implementation in Python, we run the two EnKF updating procedures discussed above with ${\mathcal M}=25$ ensemble elements for $T=5$ time
steps for different lattice sizes and monitor the computing time used in each case. The results are shown in
Figure \ref{fig:CPU_time_comparison}.
The observed computing times are shown as dots in the plot, in blue for the optimal model-based EnKF procedure and in
orange for the block updating procedure. The lines between the dots are just included to make it easier to read
the plot. As we should expect from the discussion in Section \ref{section:block_upd} we observe that the
speedup resulting from adopting the block update increases quickly with the lattice size.

\subsubsection{{Approximation error by the block update}}\label{section:lin:block_vs_opt}

When adopting the block update in an EnKF procedure we will clearly get an approximation error in each update. When assessing
this approximation error it is important to realise that the approximation error in one update may influence the behaviour
of the filter also in later iterations. First, the effect of the approximation error emerged in one update may change when later
iterations of the filter is run, the error may increase or decrease when run through later updates. Second, since the
EnKF we are using is a stochastic filter, with the stochasticity lying in the sampling of $\theta$, even a small change
after one update may have a large impact on the resulting values in later iterations. In particular
it is important to note that even if the
approximation error in one update only marginally change the distribution of the values generated in later iterations, it may have a
large impact on the actually sampled values.
In this section our focus is {first} to assess the approximation error in one update when using the block update, whereas {in the second part of this} section we concentrate on the accumulated effect of the approximations after several
iterations.


Adopting the reference time series and the data described in Section \ref{section:lin:ref_sol_and_obs} and
the assumed model and algorithmic parameters defined in Section \ref{section:num_exp_assumed_model} we isolate the
approximation error in one update by the following procedure. We first run the optimal model-based EnKF procedure
for all $T=5$ steps. Thereafter, for each $t=1,\ldots,T$, we start with the prior ensemble generated for that time step
in the optimal model-based EnKF and perform one step with the block update. One should note that since we are using the
same prior ensemble for the block update as for the optimal model-based EnKF the generated parameters $\theta$ are identical for
both updates, so the resulting difference in the updated values are only due to the difference in the $B$ matrix used
in the two procedures.

\begin{figure}
  \begin{center}
    \includegraphics[width=10cm,height=4cm]{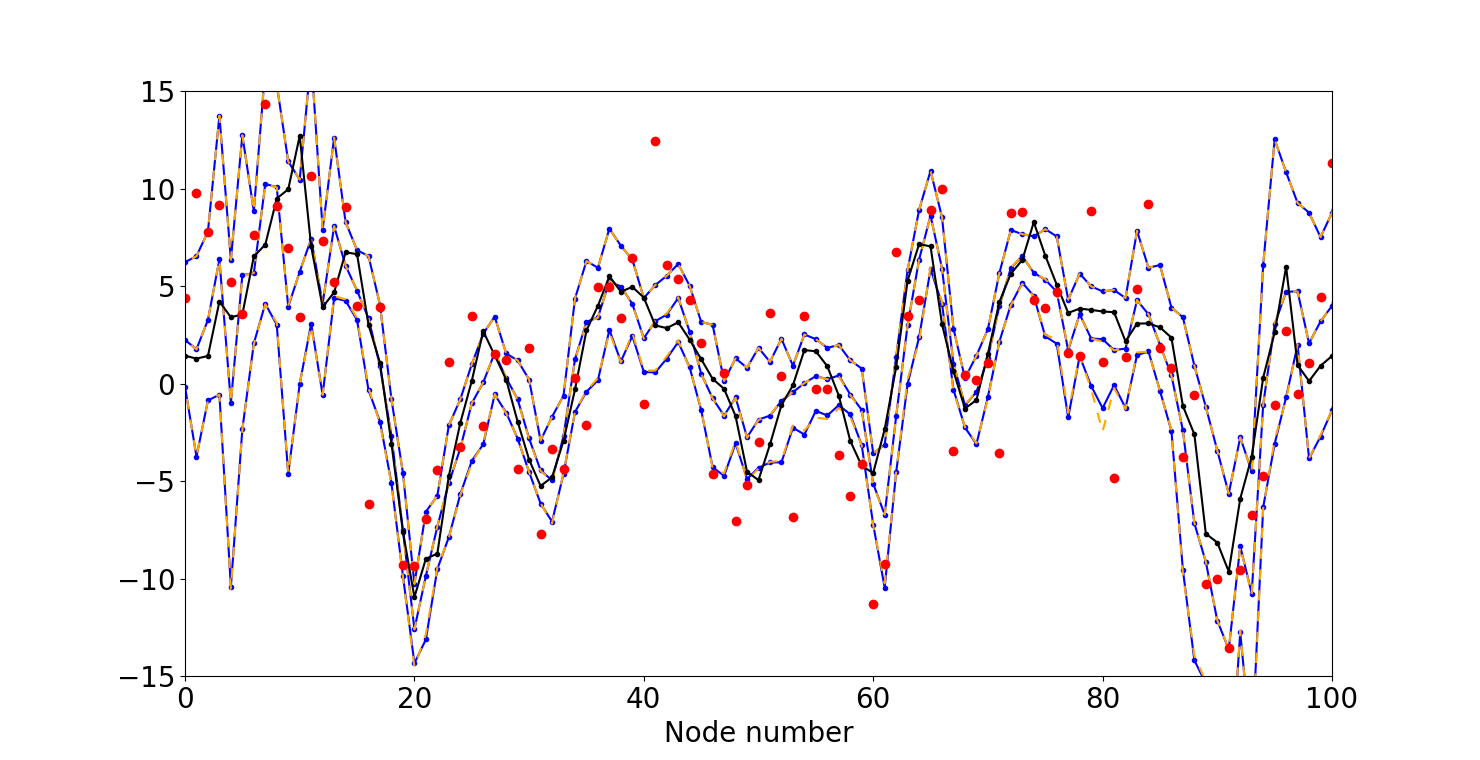}
    \caption{\label{fig:ex_1_int}{Linear example:} Assessment of the approximation error by using the block update at step $t=4$. The
      figure shows values for nodes $(50,\ell),\ell=1,\ldots,100$. The values of the reference are shown in black and the observations at $t=4$ are shown in red.
      The ensemble averages and bounds for the empirical $90\%$ {credibility} intervals when using the optimal model-based EnKF
      are shown in blue. The corresponding ensemble averages and bounds for the empirical $90\%$ {credibility} intervals when
      using the block update are shown in dashed orange}
  \end{center}
\end{figure}

Figure \ref{fig:ex_1_int}
shows results for the nodes $(50,\ell),\ell=1,\ldots,100$ at time step $t=4$. The ensemble averages and bounds of the
empirical $90\%$ {credibility} intervals for the optimal model-based EnKF procedure are shown in blue, and the
corresponding values when using the block update procedure for time $t=4$ are shown in dashed orange. For comparison
the reference is shown in black. One can observe that for most nodes the mean value and interval bounds when using
the block update are visually indistinguishable from the corresponding values when using the block update. The largest
difference can be observed for the lower bound of the {credibility} interval for $\ell=80$. Remembering that we are using
blocks of $20\times 20$ nodes in the block update it should not come as a surprise that the largest errors are for
values of $\ell$ that are multiples of $20$.

\begin{figure}
  \begin{center}
    \begin{tabular}{ccc}
      \includegraphics[width=5.1cm, height=4.0cm]{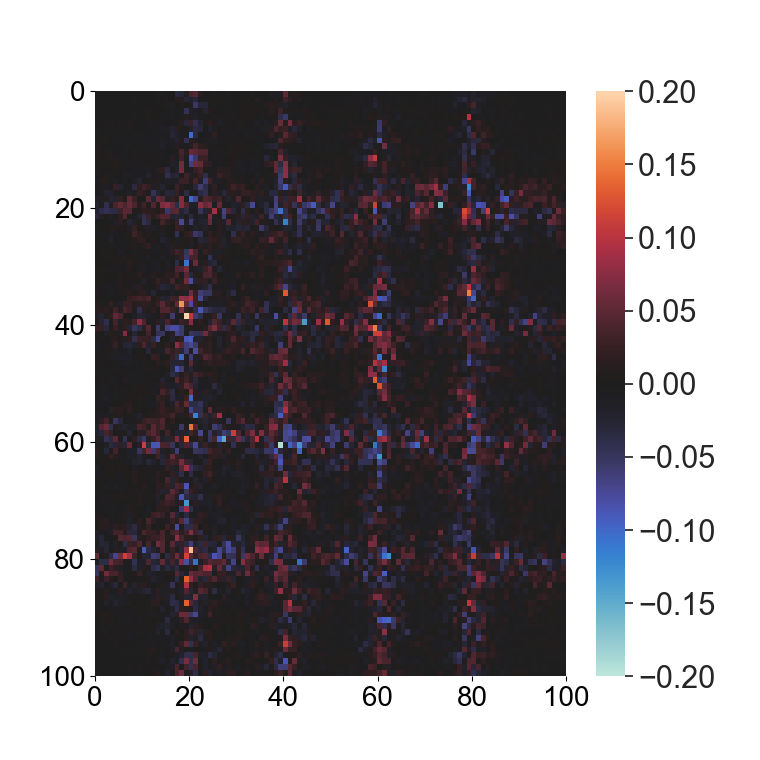}
      & ~~
      &
      \includegraphics[width=5.1cm, height=4.0cm]{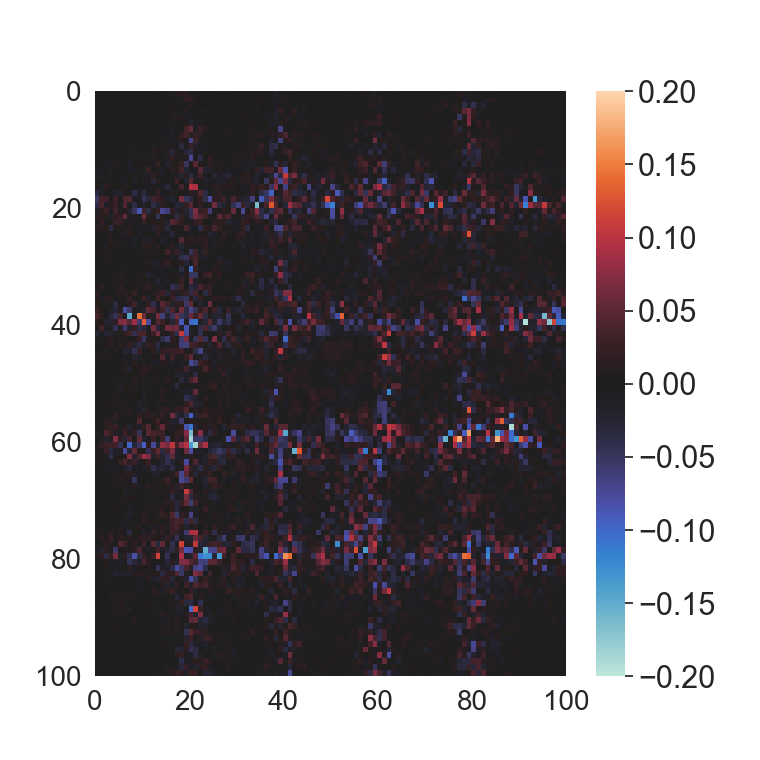}\\[-0.3cm]
      $t=1$ && $t=2$ \\
      \includegraphics[width=5.1cm, height=4.0cm]{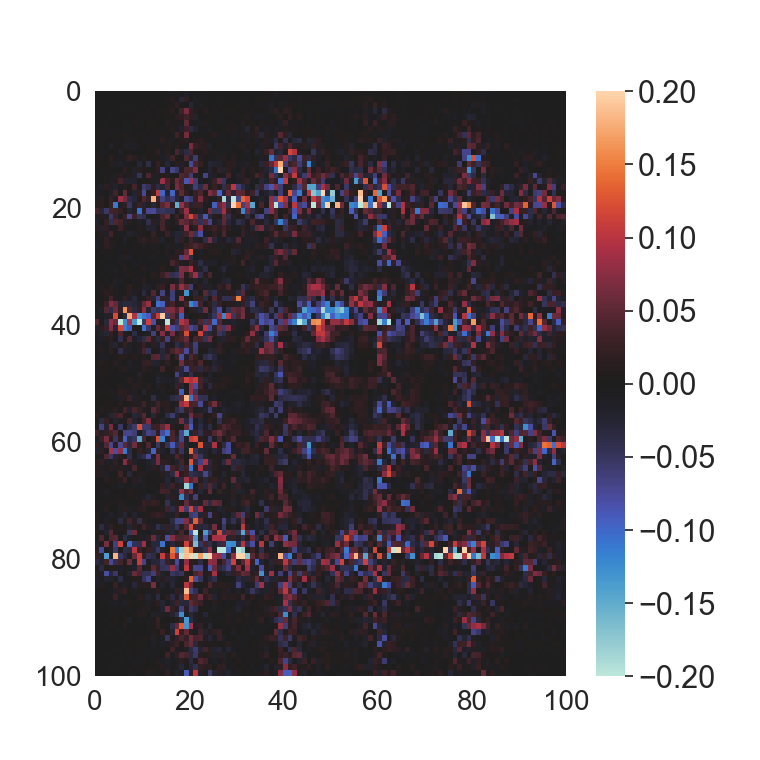}
      &&
      \includegraphics[width=5.1cm, height=4.0cm]{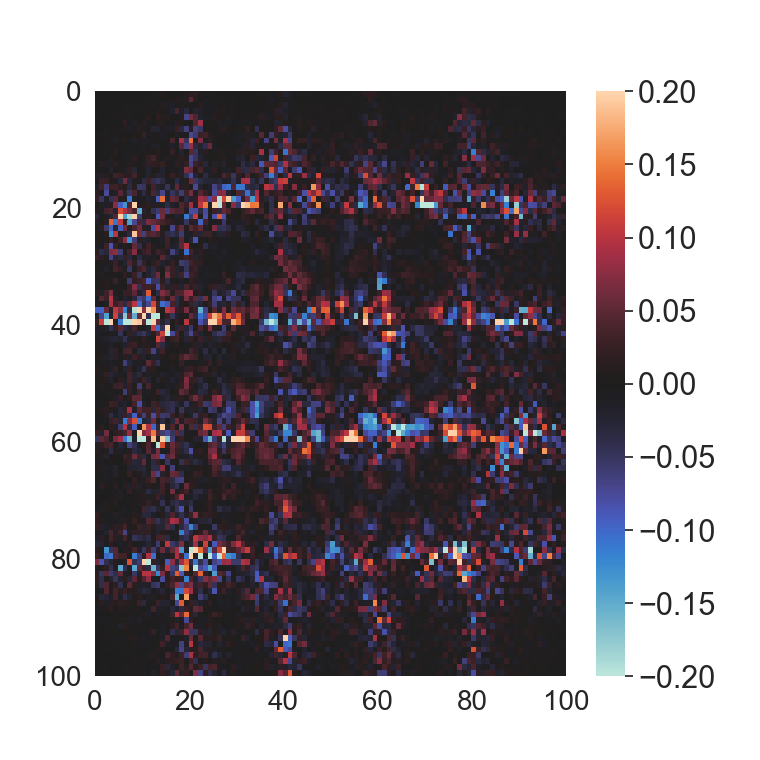}\\[-0.3cm]
      $t=3$ && $t=4$ \\
      \multicolumn{3}{c}{\begin{tabular}{c}
          \includegraphics[width=5.1cm, height=4.0cm]{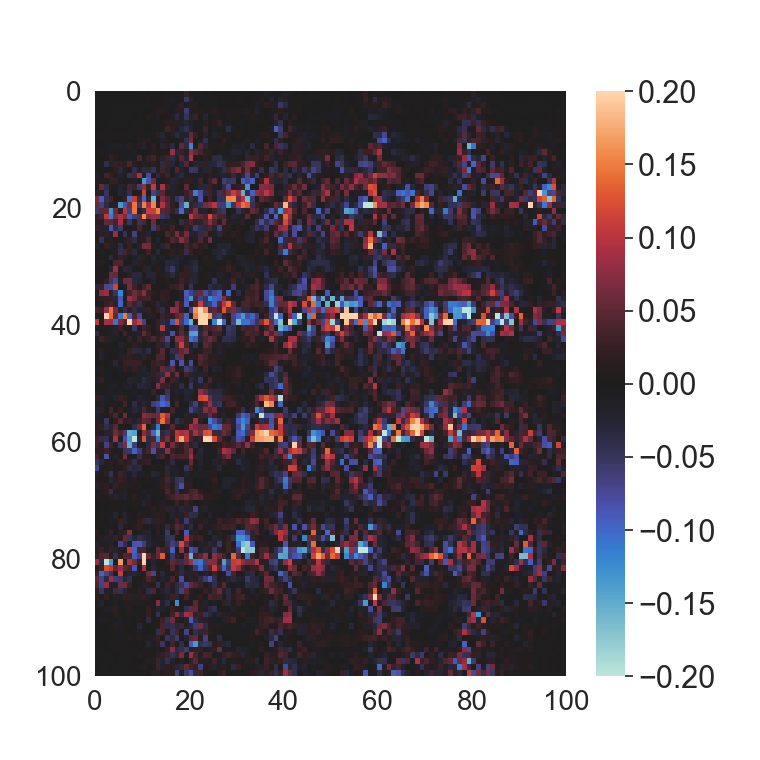}\\[-0.3cm]
          $t=5$
      \end{tabular}}
    \end{tabular}
  \end{center}
  \caption{{Linear example:} Difference between the averages of the ensembles when using the optimal model-based EnKF and
    when using the block update. Results are shown for each $t=1, \dots , 5$}
  \label{fig:ex_1}
\end{figure}

To study the approximation errors in more detail and in particular how the block structure used in
the block update influence the approximation, we in Figure \ref{fig:ex_1}
show, for each $t=1,\ldots,5$, the difference between the averages of the ensembles when using the optimal
model-based EnKF and when using the block update procedure. As one should expect we see that the largest
differences occur along the borders of the $20\times 20$ blocks used in the block update. We can, however,
also observe that the magnitude of the errors are all small compared with the spatial variability of the
underlying $x_t$ process, which explains why the approximation errors are almost invisible in Figure
\ref{fig:ex_1_int}.

\begin{figure}
  \begin{center}
    \begin{tabular}{ccc}
      \includegraphics[width=5.1cm, height=4.0cm]{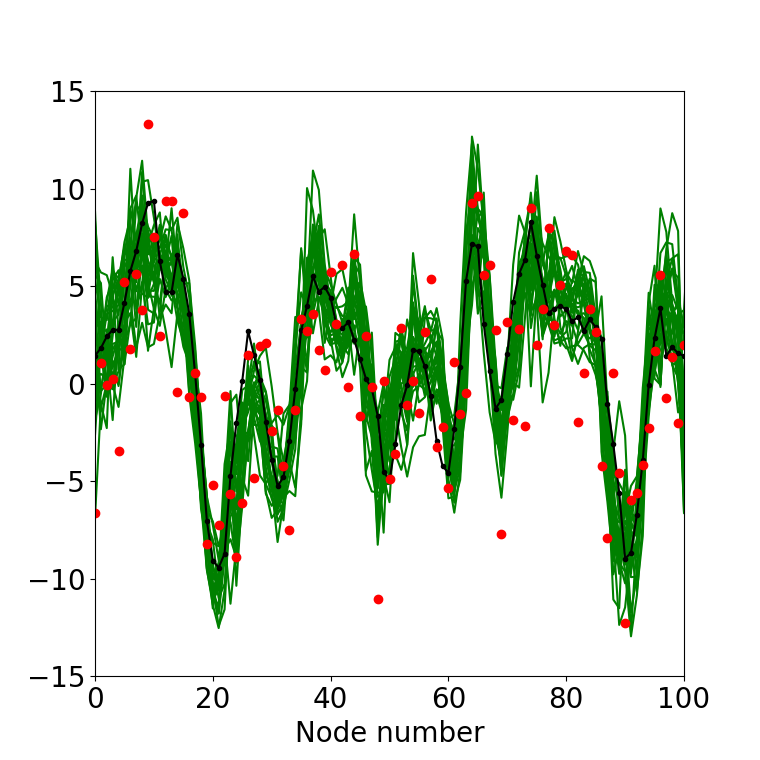}
      & ~~
      &
      \includegraphics[width=5.1cm, height=4.0cm]{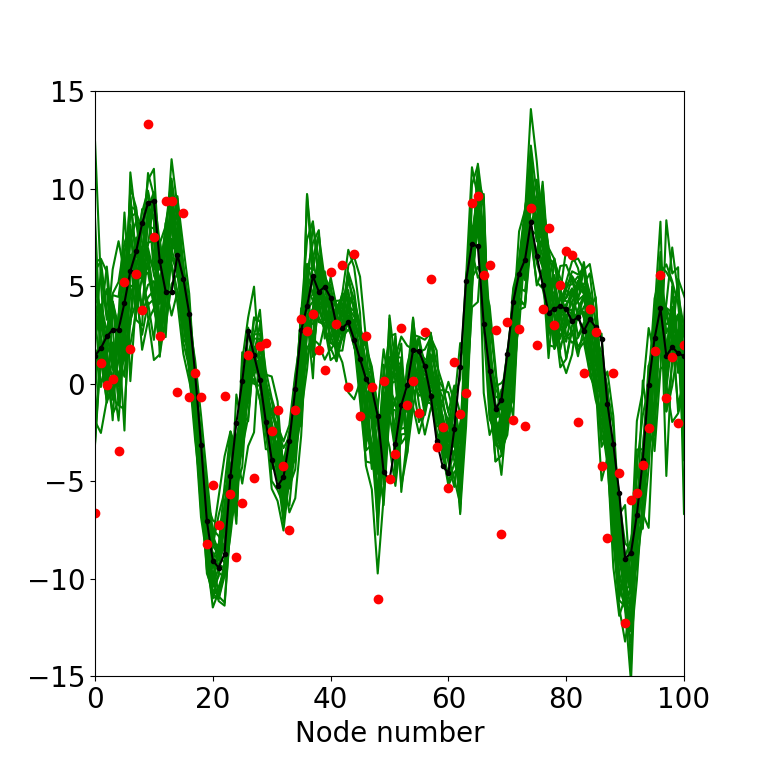} \\[-0.2cm]
      (a) & & (b) \\[-0.1cm]
      \includegraphics[width=5.1cm, height=4.0cm]{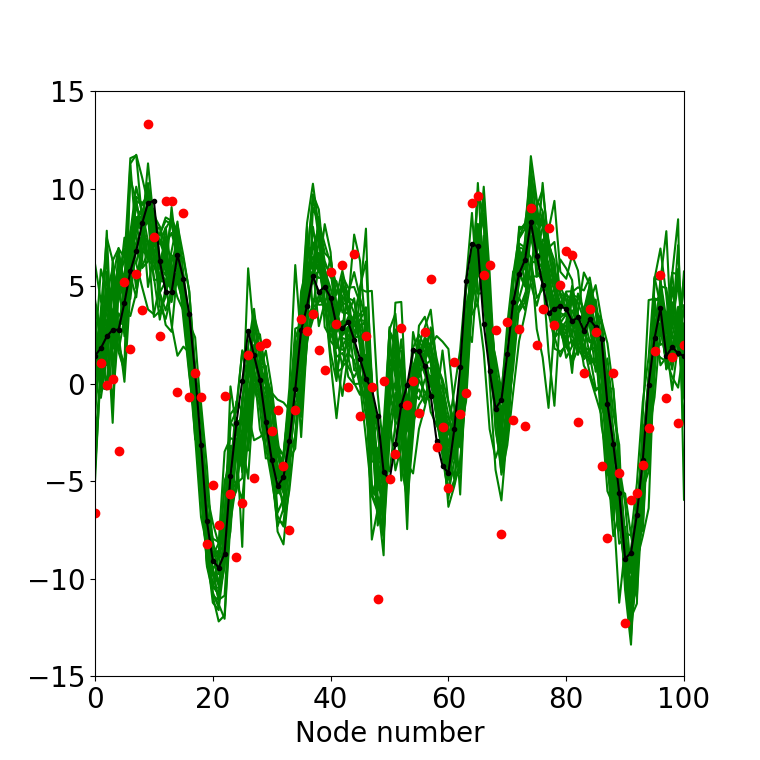}
      & &
      \includegraphics[width=5.1cm, height=4.0cm]{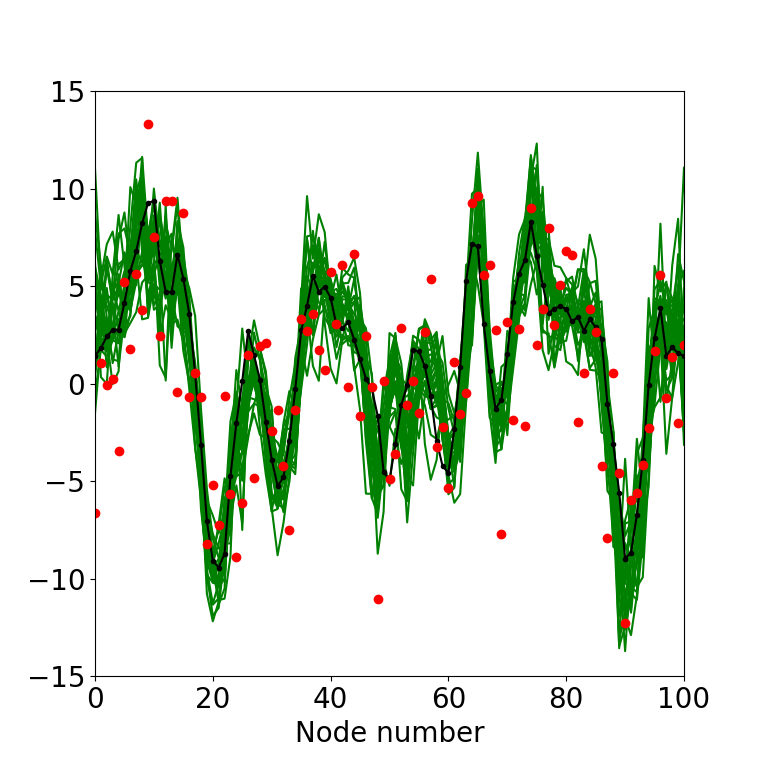} \\[-0.2cm]
      (c) && (d)
    \end{tabular}
  \end{center}
  \caption{{Linear example:} The reference solution (black) and the observations (red) along one cross section of the grid at time $T=5$. The green lines are the ensemble elements along the same cross section. Figures (a) and (c) show two simulations from the optimal update procedure, while (b) and (d) display two simulations from the block update}
  \label{fig:ex_2}
\end{figure}

{We then turn our focus to}
the accumulated effect of the approximation error over
several iterations. To do this we again adopt the reference time series and the data described in Section 
\ref{section:lin:ref_sol_and_obs}, and the assumed model and algorithmic parameters defined in Section
\ref{section:num_exp_assumed_model}. Based on this we run each of the optimal model-based EnKF and the block
update procedures several times, each time with a different initial ensemble and using different random numbers in
the updates. For two runs with each of the two 
filters, Figure \ref{fig:ex_2}
shows in green the values of all the ${\mathcal M}=25$ ensemble elements in nodes $(50,\ell),\ell=1,\ldots,100$ at
time $T=5$. The reference state at $T=5$ is shown in black and the observations at time $T=5$ are again shown as
red dots. Figures \ref{fig:ex_2}(a) and (c) show results for the optimal model-based EnKF, whereas Figures
\ref{fig:ex_2}(b) and (d) show results for the proposed block update procedure. Since all four runs are based on the
same observations they obviously have many similarities. However, all the ensembles are different. Ensembles
updated with the same updating procedure differ because of Monte Carlo variability, i.e. because they used
different initial ensembles and different random numbers in the updates. When comparing an ensemble updated
with the optimal model-based EnKF procedure with an ensemble updated with the block update, they differ both because of
Monte Carlo variability and because of the accumulated effect of the approximation errors up to time $T$. In
Figure \ref{fig:ex_2}, however, the differences between the ensembles in (a) and (b) and between the
ensembles in (c) and (d) seem similar to the differences between the ensembles in (a) and (c) and between the ensembles
in (b) and (d). So this indicates that the accumulated effect of the approximation error is small compared to the
Monte Carlo variability.

\begin{figure}
  \begin{center}
    \begin{tabular}{ccc}
        \begin{tikzpicture}[scale=0.45]
            \begin{axis}[ybar=0.5pt, width=\textwidth, ymin=0, xmin=0, xmax=1, ymax=4, bar width = 1cm, tick label style={font=\Large}]
                    
                 \addplot [color=black, fill=blue, bar width=0.005\textwidth, opacity=1] table[x=bins, y=y]  {Figures/Result_plots/Example_2/Histogram/opt_opt_stats_t=1.txt};     
                
                 \addplot [color=black, fill=red, mark=none, bar width=0.005\textwidth, opacity=0.9] table[x=bins, y=y] {Figures/Result_plots/Example_2/Histogram/opt_block_stats_t=1.txt};
            \end{axis}
    \end{tikzpicture}
    \label{fig:hist_ex_2_1}
      & ~~
      &
     \begin{tikzpicture}[scale=0.45]
             \begin{axis}[ybar=0.5pt, width=\textwidth, ymin=0, xmin=0, xmax=1, ymax=4, bar width = 1cm, tick label style={font=\Large}]
                    
                 \addplot [color=black, fill=blue, bar width=0.005\textwidth, opacity=1] table[x=bins, y=y]  {Figures/Result_plots/Example_2/Histogram/opt_opt_stats_t=2.txt};     
                
                 \addplot [color=black, fill=red, mark=none, bar width=0.005\textwidth, opacity=0.9] table[x=bins, y=y] {Figures/Result_plots/Example_2/Histogram/opt_block_stats_t=2.txt};
                
            \end{axis}
    \end{tikzpicture}\\[-0.1cm]
      $t=1$ && $t=2$ \\
      \begin{tikzpicture}[scale=0.45]
             \begin{axis}[ybar=0.5pt, width=\textwidth, ymin=0, xmin=0, xmax=1, ymax=4, bar width = 1cm, tick label style={font=\Large}]
                    
                 \addplot [color=black, fill=blue, bar width=0.005\textwidth, opacity=1] table[x=bins, y=y]  {Figures/Result_plots/Example_2/Histogram/opt_opt_stats_t=3.txt};     
                
                 \addplot [color=black, fill=red, mark=none, bar width=0.005\textwidth, opacity=0.9] table[x=bins, y=y] {Figures/Result_plots/Example_2/Histogram/opt_block_stats_t=3.txt};
                
            \end{axis}
    \end{tikzpicture}
      &&
      \begin{tikzpicture}[scale=0.45]
             \begin{axis}[ybar=0.5pt, width=\textwidth, ymin=0, xmin=0, xmax=1, ymax=4, bar width = 1cm, tick label style={font=\Large}]
                    
                 \addplot [color=black, fill=blue, bar width=0.005\textwidth, opacity=1] table[x=bins, y=y]  {Figures/Result_plots/Example_2/Histogram/opt_opt_stats_t=4.txt};     
                
                 \addplot [color=black, fill=red, mark=none, bar width=0.005\textwidth, opacity=0.9] table[x=bins, y=y] {Figures/Result_plots/Example_2/Histogram/opt_block_stats_t=4.txt};
                
            \end{axis}
    \end{tikzpicture}\\[-0.2cm]
      $t=3$ && $t=4$ \\
      \multicolumn{3}{c}{\begin{tabular}{c}
          \begin{tikzpicture}[scale=0.45]
             \begin{axis}[ybar=0.5pt, width=\textwidth, ymin=0, xmin=0, xmax=1, ymax=4, bar width = 1cm, tick label style={font=\Large}]
                    
                 \addplot [color=black, fill=blue, bar width=0.005\textwidth, opacity=1] table[x=bins, y=y]  {Figures/Result_plots/Example_2/Histogram/opt_opt_stats_t=5.txt};     
                
                 \addplot [color=black, fill=red, mark=none, bar width=0.005\textwidth, opacity=0.9] table[x=bins, y=y] {Figures/Result_plots/Example_2/Histogram/opt_block_stats_t=5.txt};
                
            \end{axis}
    \end{tikzpicture}\\[-0.2cm]
          $t=5$
      \end{tabular}}
    \end{tabular}
  \end{center}
  \caption{{Linear example:} Histograms of the two-sample Kolmogorov-Smirnov statistics. The statistics computed from two ensembles updated with the optimal ensemble are displayed in blue, while the statistics computed by two ensembles from different updating procedures are visualised in red}
  \label{fig:hist_ex_2}
\end{figure}
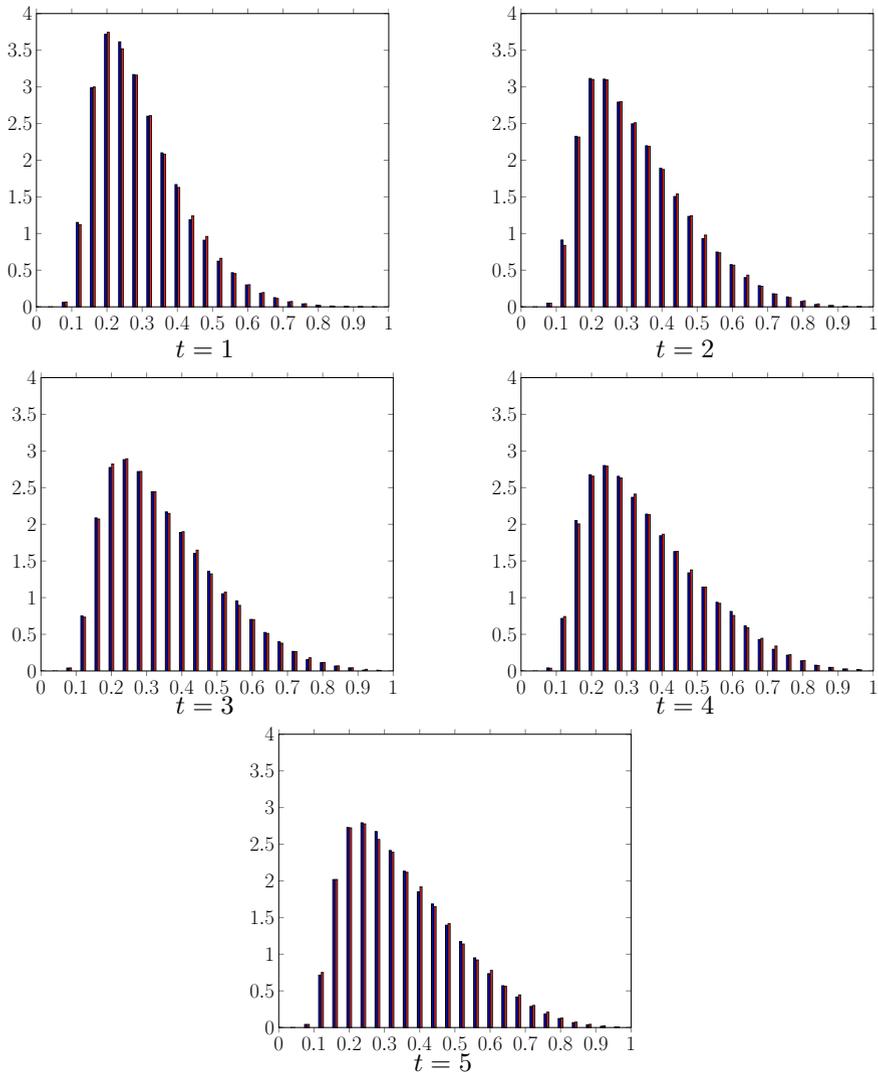

To compare the magnitude of the accumulated approximation error with the Monte Carlo variability in more detail
we adopt the two-sample
Kolmogorov-Smirnov statistic to measure the difference between the values of two ensembles in one node. More
precisely, letting
$\widehat{F}_1(x;k,\ell)$ and $\widehat{F}_2(x;k,\ell)$ denote the empirical distribution functions for the values in node
$(k,\ell)$ in two ensembles we use
\begin{equation} \label{two-sample_ks}
    D(k,\ell) = \max_x\vert \widehat{F}_1(x;k,\ell) - \widehat{F}_2(x;k,\ell)\vert 
\end{equation}
to measure the difference between the two ensembles at node $(k,\ell)$. One should note
that since we are using ${\mathcal M}=25$ elements in each ensemble the $D(k,\ell)$ is a discrete variable
with possible values $0.04d; d=0,1,\ldots,25$. Using three runs based on the optimal model-based
EnKF we compute $D(k,\ell)$ for each possible pair of ensembles, for each $t=1,\ldots,5$ and for each node
in the $100\times 100$ lattice. In Figure \ref{fig:hist_ex_2}
histogram of the resulting $30\ 000$ values for each $t=1,\ldots,5$ are shown in blue. Correspondingly we estimate
the distribution of $D(k,\ell)$ when one of the ensembles is based on the optimal model-based EnKF and the other is
using the block update, and this is shown in red in Figure \ref{fig:hist_ex_2}. Comparing the blue and the red
histograms for each $t=1,\ldots,5$ in Figure \ref{fig:hist_ex_2} one may arguably see a tendency of the mass in
the red histograms to be slightly moved to the right relative to the corresponding blue histograms. This is as
one would expect, but we also observe that the effect of the approximation error is negligible
compared to the Monte Carlo variability.

\subsubsection{{Comparison of the results with the Kalman filter output}}\label{section:block_upd_vs_KF}

{Since the forward model in this numerical example is linear and deterministic and the observation model is
  Gauss-linear, we are able to compare the block update to the Kalman filter solution.
  Figure \ref{fig:Kalman_filter_comparison}
  \begin{figure}
  \begin{center}
    \includegraphics[width=10cm,height=4cm]{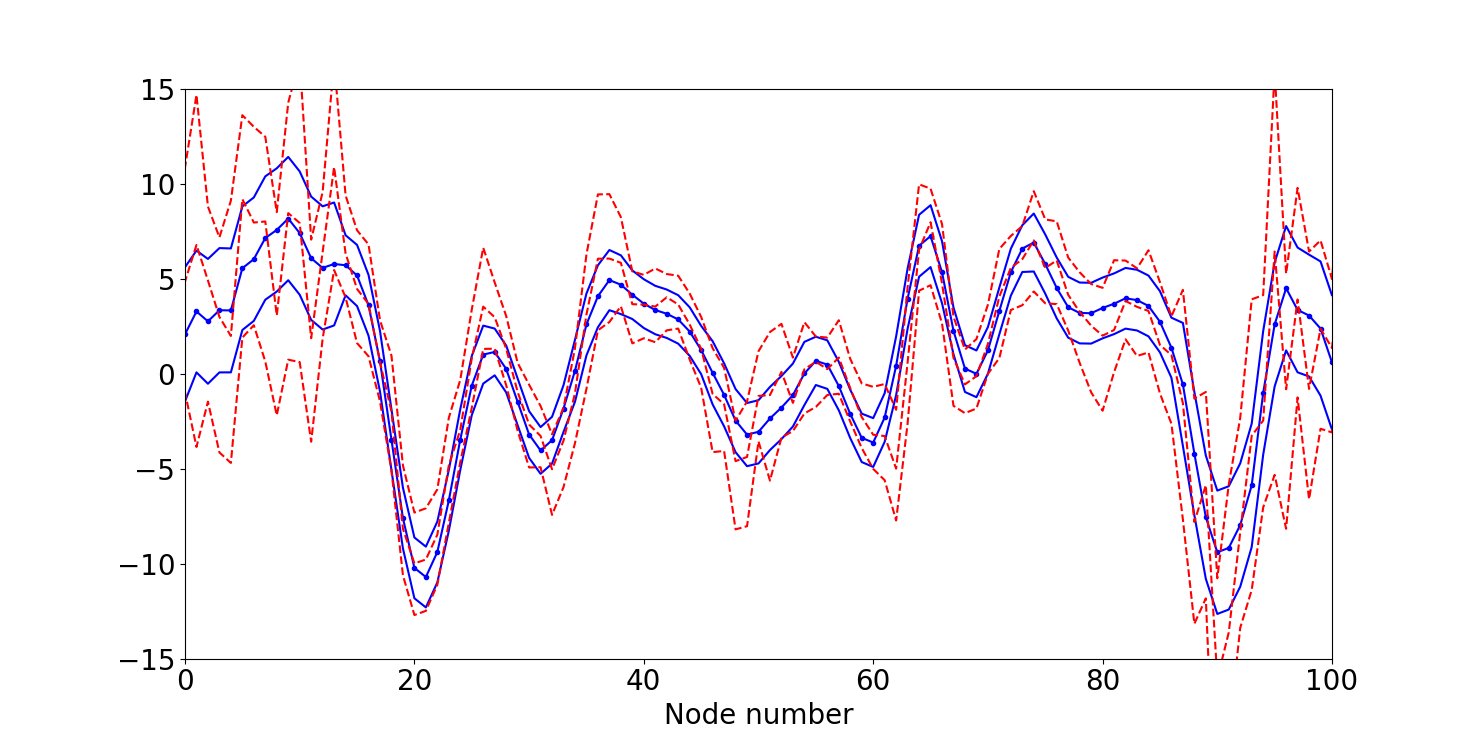}
    \caption{\label{fig:Kalman_filter_comparison}{Linear example: The blue lines display the mean of the Kalman filter solution along with the $90\%$ {credibility} intervals for one cross section at time $t=4$. The red lines visualise the empirical mean and empirical $90\%$ {credibility} interval for the block update for the same cross section at the same time step.}}
  \end{center}
\end{figure}
  displays the mean of the Kalman filter solution along with the $90\%$
  {credibility} interval for the nodes $(50,\ell), l = 1, \ldots , 100$ in blue at $t=4$. The red lines visualise the
  empirical mean and empirical $90\%$ {credibility} interval as given by one simulation of the block update.
  When comparing the two set of curves one should remember that the EnKF output is stochastic, so for another
  EnKF run a different set of curves would be produced. In general, however, we notice that the mean given by the
  block update follows the mean of the Kalman filter quite well.

  When comparing the {credibility} intervals we can observe that the intervals provided by the block update appears to
  be somewhat longer than the {credibility} intervals provided by the Kalman filter. This is as one should expect.
  At each step of the block update filter, as for any other ensemble based filter, we represent the
  {credibility} and filtering distribution
  by a rather small set of realisations. So compared to the Kalman filter, where the exact {credibility} and filtering
  distributions are represented by mean vectors and covariance matrices, we necessarily loose some information
  about the the underlying true state vector at each iteration of the block update filter, when
  representing the distributions by a set of realisations only.

  The results for times $t=1, 2, 3$ and $5$ are similar to the results for time $t=4$, except that the widths of the
  {credibility} intervals decrease with time. The shrinkage of the widths of the {credibility} intervals is clearly 
  visible for both the Kalman filter and the block update filter, but is stronger for the Kalman filter than for
  the block update filter. Such a shrinkage is as one should expect when the forward function is deterministic.
  That the shrinkage is strongest for the Kalman filter is reasonable since in each iteration of block update
  filter we loose some information about the underlying true state when just representing the distributions by
  ensembles.}

\subsection{{Non-linear example}}\label{section:non_lin_example}

{In the following we consider a non-linear example. We first describe how the reference solution and observations are generated, before we proceed to compare the results provided by the block update filter with the results of the optimal model-based EnKF.}

\subsubsection{{Reference time series and observations}}\label{section:nonlin:ref_sol_and_obs}
{The reference state for the initial time step, $x_1$, is generated in the exact same manner as for the linear case.
  The forward model used is inspired by the non-linear forward function in \citet{HEnKF}. The forward
  function is acting on each element of the state vector separately. The reference state at time
  $t>1$ for node $(k,\ell)$ is defined by
\begin{equation}
    x_t^{(k-1)\cdot s+\ell} = x_{t-1}^{(k-1)\cdot s+\ell}+0.5\cdot \arctan\left(\frac{x_{t-1}^{(k-1)\cdot s+\ell}}{2}\right).
\end{equation}
The term $0.5\cdot \arctan(x/2)$ is chosen to make the forward function clearly non-linear over the interval in which
the elements of the state vector vary. The term is displayed in Figure
\ref{fig:nonlinear_forward_func}. 
\begin{figure}
  \begin{center}
    \includegraphics[width=10.0cm,height=4cm]{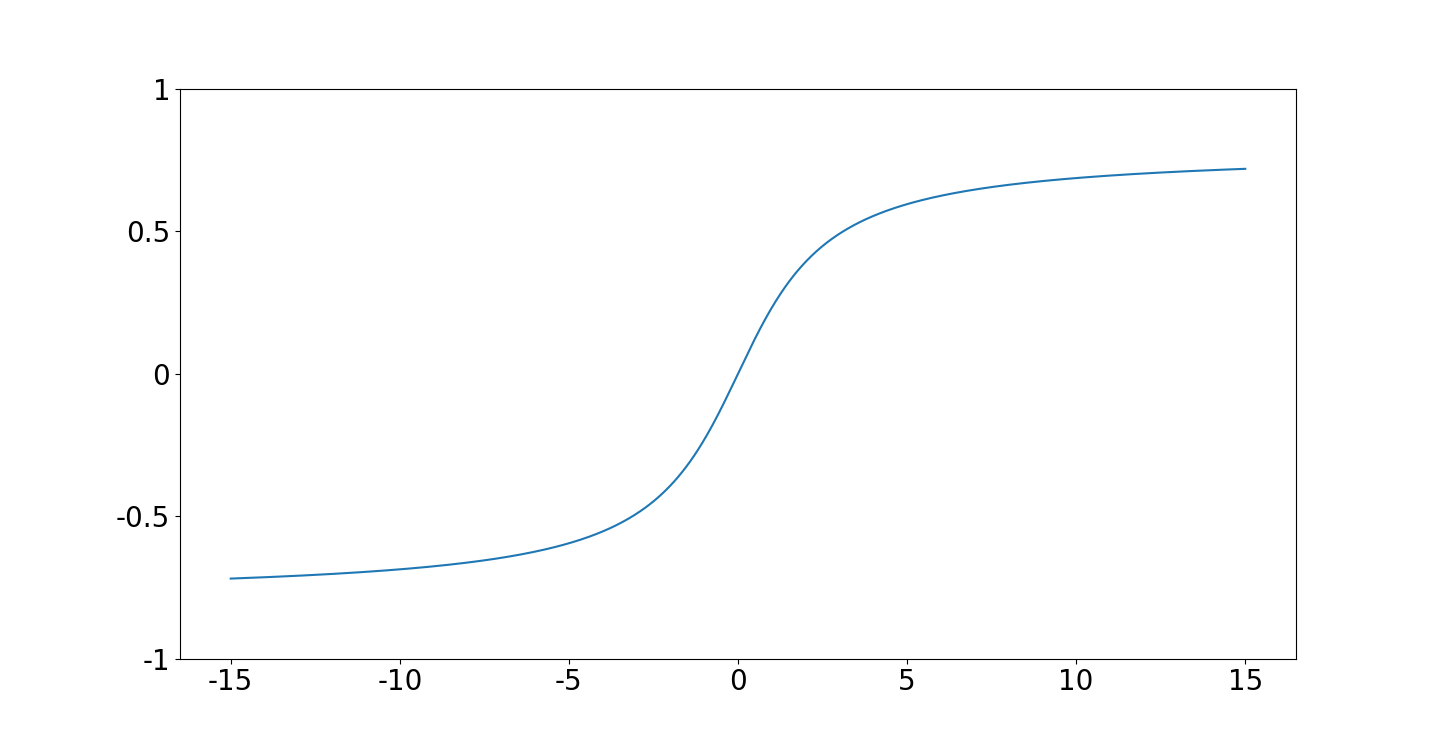}
    \caption{{Non-linear example: The term $0.5\cdot\arctan(x/2)$ in the forward function of the non-linear example}}
      \label{fig:nonlinear_forward_func} 
  \end{center}
\end{figure}
The observations are generated in the exact same manner as in the linear example. The assumed model and all the algorithmic
parameters are also chosen identical to what we used in the linear example.}

\subsubsection{{Approximation error in block update}}\label{section:nonlin:block_vs_opt}
{As for the linear example, we first consider the approximation error obtained in one update,
  and thereafter study how the approximation error accumulates over multiple iterations.

To study the effect of the approximation error in one iteration we follow the same procedure
as used in Section \ref{section:lin:block_vs_opt}. In Figure \ref{fig:ex_1_int_nonlin}
\begin{figure}
  \begin{center}
    \includegraphics[width=10cm,height=4cm]{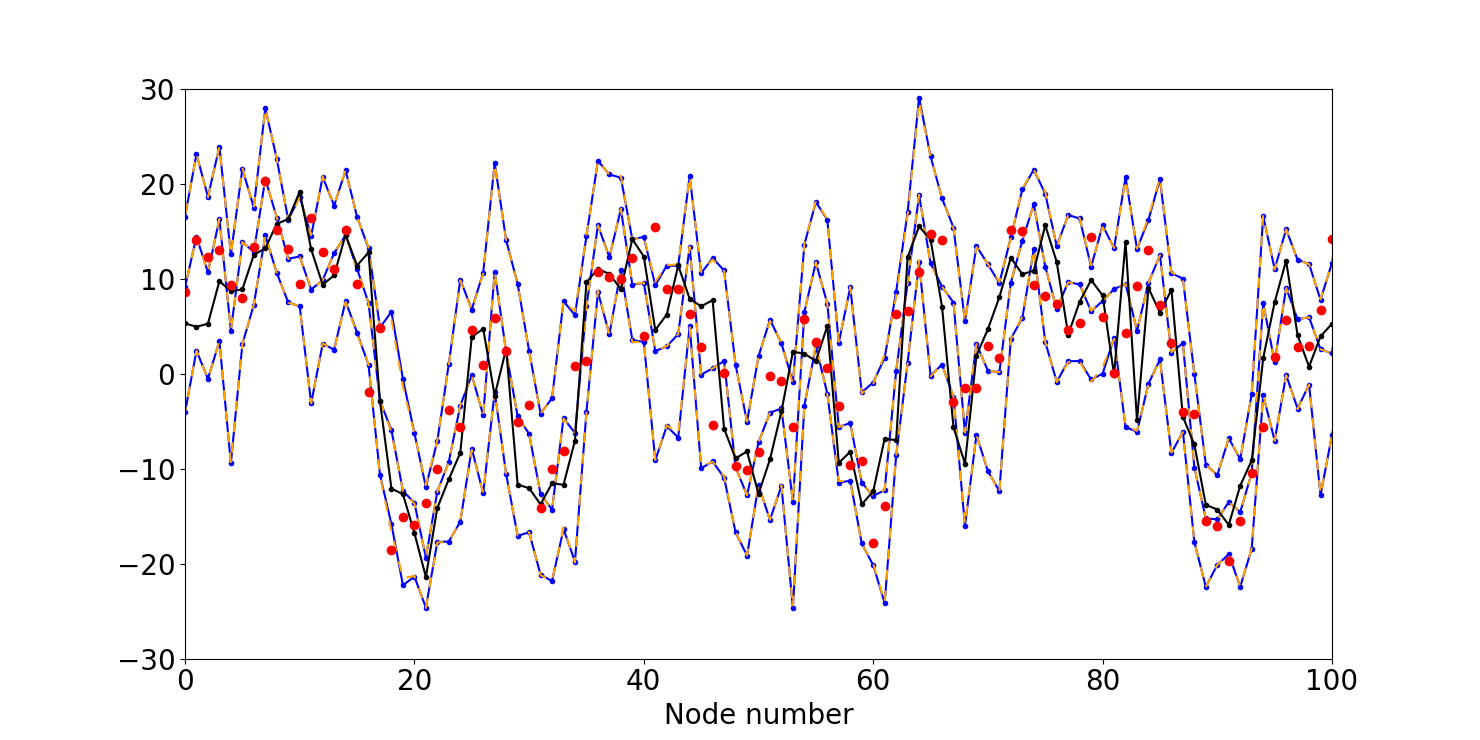}
    \caption{\label{fig:ex_1_int_nonlin}{Non-linear example: Assessment of the approximation error by using the block update at step $t=4$. The figure shows values for nodes $(50,\ell),\ell=1,\ldots,100$. The values of the reference are shown in black and the observations at $t=4$ are shown in red.
    The ensemble averages and bounds for the empirical $90\%$ {credibility} intervals when using the optimal model-based EnKF are shown in blue. The corresponding ensemble averages and bounds for the empirical $90\%$ {credibility} intervals when
      using the block update are shown in dashed orange}}
  \end{center}
\end{figure}
we compare the {credibility} intervals provided by the optimal update and block update at time $t=4$,
similar to Figure \ref{fig:ex_1_int} for the linear case. The {credibility} interval for the nodes $(50,\ell)$, $\ell=1, \ldots , 100$ provided by the optimal update is displayed in blue, while the {credibility} interval provided by the block update is visualised in orange. We notice that the {credibility} intervals provided by the two update procedures are practically speaking visually indistinguishable,
similar to what we observe for the linear case. We have also studied the spatial variability of the approximation errors by
inspecting plots of the difference of the means of the ensembles provided by the two update procedures,
similar to Figure \ref{fig:ex_1}. The results for the non-linear example are similar to what we observed in the linear case.
The largest approximation errors are again concentrated around the block borders, and the relative approximation errors
are similar to what we had with the linear forward function.


To study how the approximation errors accumulated over several iterations, we followed the same procedure as in the linear example.
When comparing ensembles resulting from the two filters, corresponding to what is done in Figure \ref{fig:ex_2}, the accumulated
effect of the approximation error again seem to be small compared to the Monte Carlo variability. 
Figure \ref{fig:hist_ex_2_nonlin}
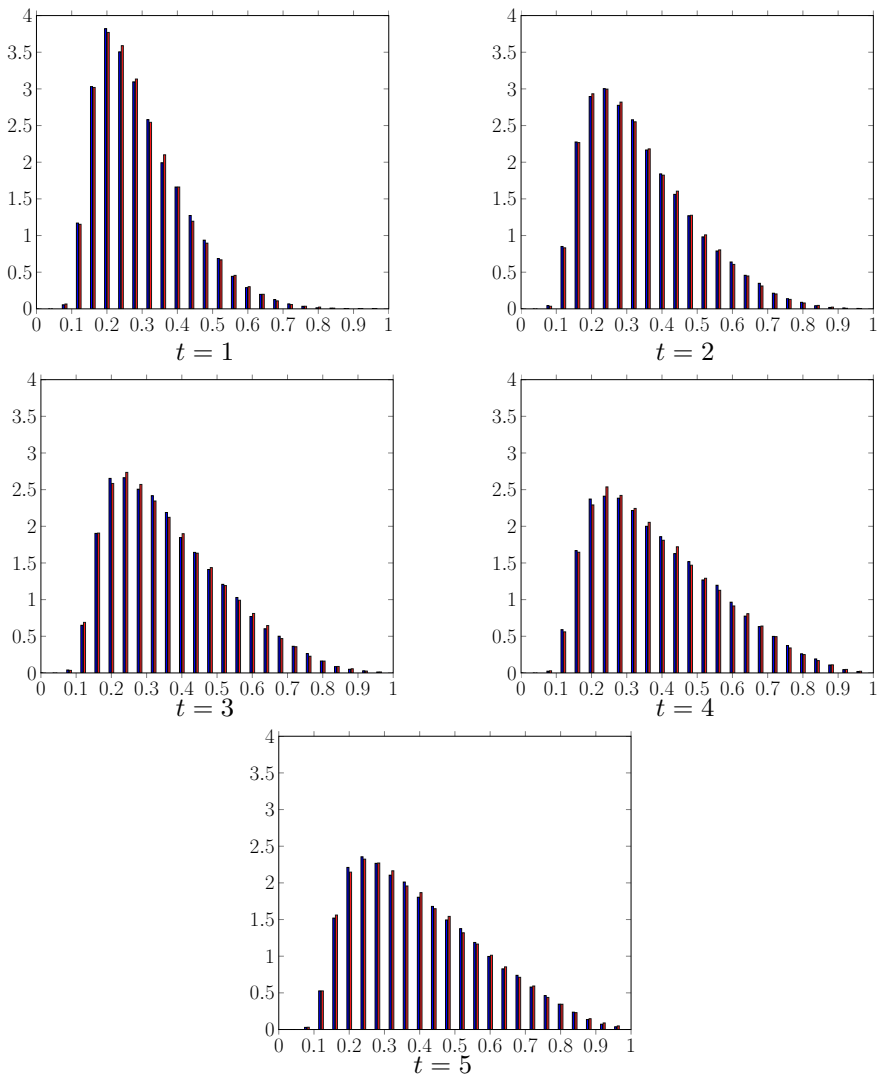
\begin{figure}
  \begin{center}
    \begin{tabular}{ccc}
        \begin{tikzpicture}[scale=0.45]
            \begin{axis}[ybar=0.5pt, width=\textwidth, ymin=0, xmin=0, xmax=1, ymax=4, bar width = 1cm, tick label style={font=\Large},black]
                    
                 \addplot [color=black, fill=blue, bar width=0.005\textwidth, opacity=1] table[x=bins, y=y]  {Figures/Result_plots/Nonlinear/Example_2/Histogram/opt_opt_stats_t=1.txt};     
                
                 \addplot [color=black, fill=red, mark=none, bar width=0.005\textwidth, opacity=0.9] table[x=bins, y=y] {Figures/Result_plots/Nonlinear/Example_2/Histogram/opt_block_stats_t=1.txt};
            \end{axis}
    \end{tikzpicture}
    \label{fig:hist_ex_2_1_nonlin}
      & ~~
      &
     \begin{tikzpicture}[scale=0.45]
             \begin{axis}[ybar=0.5pt, width=\textwidth, ymin=0, xmin=0, xmax=1, ymax=4, bar width = 1cm, tick label style={font=\Large},black]
                    
                 \addplot [color=black, fill=blue, bar width=0.005\textwidth, opacity=1] table[x=bins, y=y]  {Figures/Result_plots/Nonlinear/Example_2/Histogram/opt_opt_stats_t=2.txt};     
                
                 \addplot [color=black, fill=red, mark=none, bar width=0.005\textwidth, opacity=0.9] table[x=bins, y=y] {Figures/Result_plots/Nonlinear/Example_2/Histogram/opt_block_stats_t=2.txt};
                
            \end{axis}
    \end{tikzpicture}\\[-0.1cm]
      $t=1$ && $t=2$ \\
      \begin{tikzpicture}[scale=0.45]
             \begin{axis}[ybar=0.5pt, width=\textwidth, ymin=0, xmin=0, xmax=1, ymax=4, bar width = 1cm, tick label style={font=\Large},black]
                    
                 \addplot [color=black, fill=blue, bar width=0.005\textwidth, opacity=1] table[x=bins, y=y]  {Figures/Result_plots/Nonlinear/Example_2/Histogram/opt_opt_stats_t=3.txt};     
                
                 \addplot [color=black, fill=red, mark=none, bar width=0.005\textwidth, opacity=0.9] table[x=bins, y=y] {Figures/Result_plots/Nonlinear/Example_2/Histogram/opt_block_stats_t=3.txt};
                
            \end{axis}
    \end{tikzpicture}
      &&
      \begin{tikzpicture}[scale=0.45]
             \begin{axis}[ybar=0.5pt, width=\textwidth, ymin=0, xmin=0, xmax=1, ymax=4, bar width = 1cm, tick label style={font=\Large},black]
                    
                 \addplot [color=black, fill=blue, bar width=0.005\textwidth, opacity=1] table[x=bins, y=y]  {Figures/Result_plots/Nonlinear/Example_2/Histogram/opt_opt_stats_t=4.txt};     
                
                 \addplot [color=black, fill=red, mark=none, bar width=0.005\textwidth, opacity=0.9] table[x=bins, y=y] {Figures/Result_plots/Nonlinear/Example_2/Histogram/opt_block_stats_t=4.txt};
                
            \end{axis}
    \end{tikzpicture}\\[-0.2cm]
      $t=3$ && $t=4$ \\
      \multicolumn{3}{c}{\begin{tabular}{c}
          \begin{tikzpicture}[scale=0.45]
             \begin{axis}[ybar=0.5pt, width=\textwidth, ymin=0, xmin=0, xmax=1, ymax=4, bar width = 1cm, tick label style={font=\Large},black]
                    
                 \addplot [color=black, fill=blue, bar width=0.005\textwidth, opacity=1] table[x=bins, y=y]  {Figures/Result_plots/Nonlinear/Example_2/Histogram/opt_opt_stats_t=5.txt};     
                
                 \addplot [color=black, fill=red, mark=none, bar width=0.005\textwidth, opacity=0.9] table[x=bins, y=y] {Figures/Result_plots/Nonlinear/Example_2/Histogram/opt_block_stats_t=5.txt};
                
            \end{axis}
    \end{tikzpicture}\\[-0.2cm]
          $t=5$
      \end{tabular}}
    \end{tabular}
  \end{center}
  \caption{\label{fig:hist_ex_2_nonlin}{Non-linear example: Histograms of the two-sample Kolmogorov-Smirnov statistics. The statistics computed from two ensembles updated with the optimal ensemble are displayed in blue, while the statistics computed by two ensembles from different updating procedures are visualised in red}}
\end{figure}displays the histograms of the Kolmogorov-Smirnov statistics for $t=1, \ldots ,5$, analogous to
Figure \ref{fig:hist_ex_2} for the linear case. We arguably can observe that the tendency of the mass in the red
histograms to be moved to the right compared to the corresponding blue histograms, is slightly stronger in this
non-linear example than in the linear case.}

\section{Closing remarks} \label{section:closing_remarks}
In this paper we propose two changes in the model-based EnKF procedure introduced in \citet{LoeTjelmeland2021}. Our
motivation is to get a procedure that is computationally faster than the original one, so that it is feasible to use
it also in situations with high dimensional state vectors. The first change we introduce is to formulate the assumed model
in terms of precision matrices instead of covariance matrices, and to adopt a prior for the precision matrix that ensures
the sampled precision matrices to be sparse. The second change we propose is to adopt the block update, which allows us to
do singular value decompositions of many smaller matrices instead of for one large one. 

In a simulation example we have studied both the computational speedup and the associated approximation error resulting
when adopting the proposed procedure. The computational speedup is substantial for high dimensional state vectors
and this allows the proposed filter to be run on
much larger problems than can be done with the original formulation. At the same time the approximation error
resulting from using the introduced block updating is negligible compared to the Monte Carlo variability inherent in both
the original and the proposed procedures.

In order to further investigate the strengths and weaknesses of the proposed approach, it should be applied on more examples.
It is of interest to gain more experience with the proposed procedure both in other simulation examples and in real data
situations. In particular it is of interest to experiment more with different sizes for the $C_b$, $D_b$ and $E_b$ blocks to
try to find the best sizes for these when taking both the computational time and the approximation quality into account.
As for all EnKFs the proposed procedure is ideal for parallel computation and with an implementation more tailored for
parallellisation it should be possible get a code that is running much faster than our more basic implementation.

	\backmatter
	
	\bmhead{Funding} 
	
	Geophysics and Applied Mathematics in Exploration and Safe production (GAMES) at NTNU (Research Council of Norway; Grant No. 294404).
	
	\section*{Declarations}
	
	\bmhead{Conflict of interest} 
	No conflicts of interest.

\section*{Appendices}
\appendix

\section{Derivation of \texorpdfstring{$\theta=(\mu,Q)$}{} from \texorpdfstring{$\phi$}{} and \texorpdfstring{$\eta$}{}}\label{section:derivation_prec_mx}

To find how $\theta=(\mu,Q)$ is given from $\phi$ and $\eta$ we restrict the Gaussian densities
$f(x\vert \theta)$ and $f(x\vert \phi,\eta)$ to be identical. The density $f(x\vert \theta)$ is given by
\begin{equation}\label{eq:xGivenTheta}
\begin{aligned}
  f(x&\vert \theta) \propto \exp\left\{-\frac{1}{2}(x-\mu)^TQ(x-\mu)\right\} = \exp\left\{ - \frac{1}{2}x^TQx + \mu^TQx\right\}\\
    &= \exp\left\{\sum_{k=1}^{n_x}\left(-\frac{1}{2}Q^{k,k}\right)(x^k)^2 + \sum_{\ell < k} \left(-Q^{k,\ell}\right)x^kx^{\ell} +
    \sum_{k=1}^{n_x}\left(\sum_{\ell=1}^{n_x}\mu^{\ell}Q^{\ell,k}\right) x^k\right\}.
\end{aligned}
\end{equation}
Using the definition of $f(x\vert \phi,\eta)$ from Section \ref{section:new_prior} we have
\begin{equation}\label{eq:xGivenPhiEta}
  \begin{aligned}
    f(x\vert \phi,\eta) &= \prod_{k=1}^{n_x} f(x^k\vert x^{\Lambda_k},\eta^k,\phi^k)\\
    &\propto \prod_{k=1}^{n_x} \exp\left\{-\frac{1}{2\phi^k}(x^k-(1,(x^{\Lambda_k})^T)\cdot \eta^k)^2\right\}\\
    &= \exp\left\{ -\frac{1}{2}\sum_{k=1}^{n_x} \frac{1}{\phi^k}\left(x^k - \eta^{k,1}-\sum_{\ell=2}^{\vert \Lambda_k\vert }
    \eta^{k,\ell} x^{\Lambda_k(\ell)}\right)^2\right\}\\
    &\propto \exp\left\{ \sum_{k=1}^{n_x} \left(-\frac{1}{2\phi^k} - \frac{1}{2}\sum_{\ell:k\in\Lambda_{\ell}}\frac{1}{\phi^{\ell}}
    \left(\eta^{\ell,\Lambda_\ell^{-1}(k)+1}\right)^2\right)(x^k)^2\right.\\
    &\hspace*{0.15cm} + \sum_{\ell < k} \left( \frac{1}{\phi^k}\eta^{k,\Lambda_k^{-1}(\ell)+1}\mathbb{I}(\ell\in\Lambda_k) -\sum_{t:k,\ell\in\Lambda_t}\frac{1}{\phi^t}
    \eta^{t,\Lambda_t^{-1}(k)+1}\eta^{t,\Lambda_t^{-1}(\ell)+1}\right) x_kx_{\ell}\\
    &\hspace*{0.15cm}\left. + \sum_{k=1}^{n_x} \left(\frac{\eta^{k,1}}{\phi^k} - \sum_{\ell:k\in\Lambda_{\ell}} \frac{1}{\phi^{\ell}}
    \eta^{\ell, 1}\eta^{\ell,\Lambda_{\ell}^{-1}(k)+1}\right) x^k
    \right\},
  \end{aligned}
\end{equation}
where $\Lambda^{-1}_{\ell}(k)$ is the inverse function of $\Lambda_{\ell}(k)$, i.e. $\Lambda_{\ell}^{-1}(k)=t\Leftrightarrow
\Lambda_{\ell}(t)=k$, and $\mathbb{I}(\ell \in \Lambda_k)$ is the indicator function returning one if 
$\ell\in\Lambda_k$ and zero otherwise. By setting equal corresponding coefficients in front of $(x^k)^2$, $x_kx_{\ell}$
and $x_k$ in (\ref{eq:xGivenTheta})
and (\ref{eq:xGivenPhiEta}) we can identify the elements of $\mu$ and $Q$. Setting equal the coefficients in front of
$(x^k)^2$ we get the diagonal elements of $Q$,
\begin{equation}\label{Qkk}
  Q^{k,k} = \frac{1}{\phi^{k}} + \sum_{\ell:k\in\Lambda_{\ell}}\frac{1}{\phi^\ell}\left(\eta^{\ell,\Lambda_{\ell}^{-1}(k)+1}\right)^2.
\end{equation}
To get the non-diagonal elements we set equal the coefficients in front of $x_kx_{\ell}$. For $\ell<k$ we have
\begin{equation}\label{Q_app}
  Q_{k,\ell} = Q_{\ell,k} = - \frac{1}{\phi^k}\eta^{k,\Lambda_k^{-1}(\ell)+1}\mathbb{I}(\ell\in\Lambda_k) +\sum_{t:k,\ell\in\Lambda_t}\frac{1}{\phi^t}
    \eta^{t,\Lambda_t^{-1}(k)+1}\eta^{t,\Lambda_t^{-1}(\ell)+1}.
\end{equation}
In particular one should note that for $\ell<k$ we get $Q_{k,\ell}=0$ if $\ell\not\in\Lambda_k$ and there is no node $t$ so that
$k,\ell\in\Lambda_t$. {Thus, if the sequential neighbourhoods are chosen to be sufficiently small, $Q$
  will be sparse. For a more detailed discussion see Appendix \ref{section:seq_neigh_sparsity_Q_app}.} Finally, by equating corresponding coefficients in front of $x_k$
we get
\begin{equation}
  \sum_{\ell=1}^{n_x}\mu^{\ell}Q^{\ell,k} = \frac{\eta^{k,1}}{\phi^k} - \sum_{\ell:k\in\Lambda_{\ell}} \frac{1}{\phi^{\ell}}
    \eta^{\ell, 1}\eta^{\ell,\Lambda_{\ell}^{-1}(k)+1},
\end{equation}
which can be used to find the elements of the mean vector $\mu$. Since $Q$ is sparse the strategy discussed in Section
\ref{section:num_prop_sparse_matrix} can be used to compute $\mu$ efficiently.

\section{Sparseness of $Q$}\label{section:seq_neigh_sparsity_Q_app}
{In the following we discuss the relationship between the sequential neighbourhood and the structure of
  the precision matrix $Q$. We first identify a general expression for the set of non-zero elements in $Q$, and
  thereafter discuss the situation in more detail when the state vector is associated with the nodes in a
  two-dimensional lattice.

  From (\ref{Qkk}) and (\ref{Q_app}) we see that element $(k,\ell)$ of the precision matrix, $Q_{k,\ell}$, may be non-zero only when
  \begin{equation}\label{eq:neighbourhood}
    \ell \in \partial k = \{k\} \cup \Lambda_k \cup \{ \ell\in{\cal S}:\exists t\in{\cal S} \mbox{~so that~} k,\ell\in\Lambda_t\},
  \end{equation}
  where ${\cal S}$ is the set of all nodes in the lattice. A perhaps more instructive formulation of the set
  $\partial k$ is as a union of sequential neighbourhoods,
  \begin{equation}\label{eq:partial}
    \partial k = \bigcup_{t\in{\cal S}: k\in \Lambda_t\cup\{t\}}\left( \Lambda_t\cup \{t\}\right).
  \end{equation}

  Now consider a situation where the state vector is associated with the nodes in a two-dimensional
  lattice and assume all the sequential neighbourhoods to be of the same form, except for nodes close to the
  lattice boundaries where the sequential neighbourhood is necessarily smaller. In this situation
  (\ref{eq:partial}) can be used to
  identify an upper bound on the number of non-zero elements in $Q$. The situation for this two-dimensional
  lattice case is illustrated in Figure \ref{fig:seq_neigh_app}.
  \begin{figure}
    \begin{center}
        \begin{tikzpicture}[scale=0.5]
            
            \draw[black, thick] (-0.50,-0.50) rectangle (12.50,12.50);
              
            \foreach \x in {0,1,..., 12}{
                \foreach \y in {0,1,..., 12}{
                    \node at (\x,\y)[circle, fill, inner sep = 1.5pt, black]{};
                }
            }


        \draw[decorate,decoration = {brace},black] (8.20, 8.00) -- (8.20,6.00);
        \draw[black, thick] (8.55, 7.00) node {$u$};


        \draw[decorate,decoration = {brace},black] (8.00, 5.80) -- (3.00,5.80);
        \draw[black, thick] (5.50, 5.35) node {$v$};
        

        \draw[red, thick] (2.85,5.85) rectangle (8.15,8.15);
        




        \node at (6, 6)[circle, fill, inner sep = 1.5pt, green]{};

        \node at (5, 6)[circle, fill, inner sep = 1.5pt, red]{};
        \node at (4, 6)[circle, fill, inner sep = 1.5pt, red]{};
        \node at (3, 6)[circle, fill, inner sep = 1.5pt, red]{};

        \node at (5, 7)[circle, fill, inner sep = 1.5pt, red]{};
        \node at (4, 7)[circle, fill, inner sep = 1.5pt, red]{};
        \node at (3, 7)[circle, fill, inner sep = 1.5pt, red]{};

        \node at (8, 7)[circle, fill, inner sep = 1.5pt, red]{};
        \node at (7, 7)[circle, fill, inner sep = 1.5pt, red]{};
        \node at (6, 7)[circle, fill, inner sep = 1.5pt, red]{};
        
        \node at (5, 8)[circle, fill, inner sep = 1.5pt, red]{};
        \node at (4, 8)[circle, fill, inner sep = 1.5pt, red]{};

        \node at (8, 8)[circle, fill, inner sep = 1.5pt, red]{};
        \node at (7, 8)[circle, fill, inner sep = 1.5pt, red]{};
        \node at (6, 8)[circle, fill, inner sep = 1.5pt, red]{};

         \node at (7, 6)[circle, fill, inner sep = 1.5pt, blue]{};
         \node at (8, 6)[circle, fill, inner sep = 1.5pt, blue]{};
         \node at (9, 6)[circle, fill, inner sep = 1.5pt, blue]{};

         \node at (7, 5)[circle, fill, inner sep = 1.5pt, blue]{};
         \node at (8, 5)[circle, fill, inner sep = 1.5pt, blue]{};
         \node at (9, 5)[circle, fill, inner sep = 1.5pt, blue]{};
         \node at (7, 4)[circle, fill, inner sep = 1.5pt, blue]{};
         \node at (8, 4)[circle, fill, inner sep = 1.5pt, blue]{};
         \node at (5, 5)[circle, fill, inner sep = 1.5pt, blue]{};
         \node at (6, 5)[circle, fill, inner sep = 1.5pt, blue]{};
         \node at (5, 4)[circle, fill, inner sep = 1.5pt, blue]{};
         \node at (6, 4)[circle, fill, inner sep = 1.5pt, blue]{};
         \node at (4, 4)[circle, fill, inner sep = 1.5pt, blue]{};
         \node at (4, 5)[circle, fill, inner sep = 1.5pt, blue]{};



        

        \node at (3, 5)[circle, fill, inner sep = 1.5pt, blue]{};
        \node at (3, 4)[circle, fill, inner sep = 1.5pt, blue]{};
        \node at (2, 5)[circle, fill, inner sep = 1.5pt, blue]{};
        \node at (2, 4)[circle, fill, inner sep = 1.5pt, blue]{};
        \node at (1, 5)[circle, fill, inner sep = 1.5pt, blue]{};
        \node at (1, 4)[circle, fill, inner sep = 1.5pt, blue]{};
        \node at (1, 6)[circle, fill, inner sep = 1.5pt, blue]{};
        \node at (2, 6)[circle, fill, inner sep = 1.5pt, blue]{};
        \node at (2, 7)[circle, fill, inner sep = 1.5pt, blue]{};
        \node at (10, 5)[circle, fill, inner sep = 1.5pt, blue]{};
        \node at (10, 6)[circle, fill, inner sep = 1.5pt, blue]{};
        \node at (11, 6)[circle, fill, inner sep = 1.5pt, blue]{};
        \node at (11, 7)[circle, fill, inner sep = 1.5pt, blue]{};
        \node at (10, 7)[circle, fill, inner sep = 1.5pt, blue]{};
        \node at (9, 7)[circle, fill, inner sep = 1.5pt, blue]{};
        \node at (11, 8)[circle, fill, inner sep = 1.5pt, blue]{};
        \node at (10, 8)[circle, fill, inner sep = 1.5pt, blue]{};
        \node at (9, 8)[circle, fill, inner sep = 1.5pt, blue]{};

         \draw[green, thick] (0.75, 3.75) rectangle (11.25,8.25);

        \end{tikzpicture}
    \end{center}
    
    \caption{{The green node represents node $k$, while the red nodes represent $\Lambda_k$. The nodes inside the red rectangle visualises $\Psi_k$. The blue nodes represent the remaining nodes in $\partial k$, while the nodes inside the green rectangle represent $\Omega_k$}}
    \label{fig:seq_neigh_app}
\end{figure}
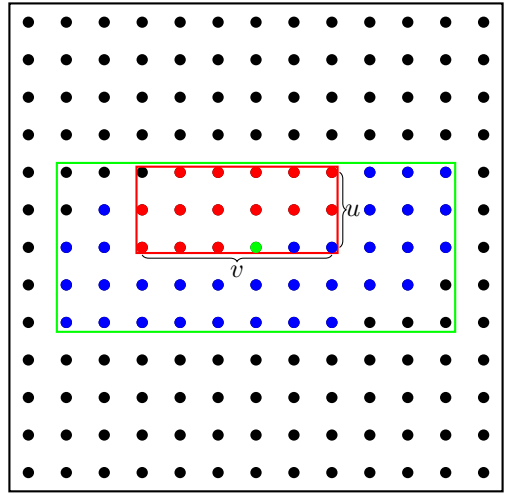
  In the figure the red nodes are sequential neighbours of the green node. Assuming the green node to represent
  element $k$ in the state vector, we let $\Psi_k$ denote the set of nodes inside the smallest rectangle that
  contains node $k$ and all the sequential neighbours of $k$. By construction we then have $\Lambda_k\cup \{k\}\subseteq\Psi_k$.
  In Figure \ref{fig:seq_neigh_app} the set $\Psi_k$ consists of all nodes inside the
  red rectangle. Letting $u$  and $v$ denote the
  vertical and horizontal dimensions, respectively, of the rectangle defining $\Psi_k$,
  again as illustrated in Figure \ref{fig:seq_neigh_app}, we get
  that the number of nodes in $\Psi_k$ is $\vert\Psi_k\vert=(u+1)(v+1)$. In Figure \ref{fig:seq_neigh_app} we
  have $u=2$ and $v=5$ and $\vert\Psi_k\vert=18$. As $\Lambda_k\cup \{k\}\subseteq\Psi_k$ it follows
  from (\ref{eq:partial}) that
  \begin{equation}
    \partial k \subseteq \Omega_k = \bigcup_{t:k\in\Psi_t}\Psi_t.
  \end{equation}
  Thereby an upper bound on the number of non-zero elements in row $k$ of $Q$ is the number of nodes in the
  set $\Omega_k$. Since $\Psi_k$ is defined by a rectangle, the set $\Omega_k$ can also be defined from a
  rectangle. In Figure \ref{fig:seq_neigh_app}, the set $\Omega_k$ consists of all nodes inside the green
  rectangle, whereas the set $\partial k$ consists of the green, red and blue nodes in the figure.
  The vertical and horizontal dimensions of the rectangle defining $\Omega_k$ become
  $2u$ and $2v$, respectively, and the number of elements in $\Omega_k$ becomes $\vert\Omega_k\vert=(2u+1)(2v+1)$.
  So the the number of non-zero elements in row $k$ of $Q$ is bounded by $(2u+1)(2v+1)$. Moreover,
  since we have $n_x$ elements in the state vector, we will have at most
  $(2u+1)(2v+1)n_x$ non-zero elements in the precision matrix $Q$.

In the two-dimensional lattice case we can also provide an upper bound for the bandwidth of $Q$.
  Still assuming that each node in the lattice is assigned similar sequential neighbourhoods,
  an upper bound for the bandwidth is the maximum distance between node $k$ and any node $\ell\in\Omega_k$, i.e. $su+v$, where $s$ is the number of columns in the grid.
}

\section{Derivation of posterior for hyperpriors}\label{section:posterior_derivation_appendix}

In the following we derive the posterior distribution for $(\eta, \phi)$. In the derivation we
simplify the notation by omitting the subscripts on the $f$'s as it should be clear
from the context what densities we are dealing with. 
We start by identifying $f(\phi\vert  x, z^{(m)})$ and thereafter find $f(\eta\vert \phi,x,z^{(m)})$.

Using the definitions of $f(x\vert  \eta, \phi)$, $f(\phi)$ and $f(\eta\vert \phi)$ given in Section
\ref{section:new_prior}, and that $z^{(m)}$ and $x$ are conditionally independent given
$\phi$ and $\eta$, we get
\begin{align*}
  f(\phi\vert z^{(m)},x)&\propto f(\phi)f(z^{(m)},x\vert \phi) \\
  &= f(\phi) \int f(\eta\vert \phi) f(z^{(m)},x\vert \phi, \eta) d\eta    
\\&= f(\phi) \int f(\eta\vert \phi) f(x\vert \phi, \eta) f(z^{(m)}\vert \phi, \eta) d\eta   
\\&= f(\phi) \int f(\eta\vert \phi) f(x\vert \phi, \eta) \prod_{i\neq m} f(x^{(i)}\vert \phi, \eta) d\eta   
\\&= \prod_{k=1}^K \left[  f(\phi^k) \int f(\eta^k\vert \phi^k) f(x^k\vert \phi^k, \eta^k, x^{\Lambda_k}) \prod_{i\neq m} f(x^{(i),k}\vert \phi^k, \eta^k, x^{\Lambda_k}) d\eta^k \right]
\\&\propto \prod_{k=1}^K f(\phi^k\vert z^{(m)},x),
\end{align*}
where
\begin{equation}
\begin{aligned} \label{appendix_phi_marginal_derivation}
  f(\phi^k&\vert z^{(m)},x) \propto f(\phi^k) \int  f(\eta^k\vert \phi^k) f(x^k\vert \phi^k, \eta^k, x^{\Lambda_k}) \prod_{i \neq m}
  f(x^{(i),k}\vert \phi^k, \eta^k, x^{\Lambda_k}) d\eta^k \\
    &= f(\phi^k)\cdot \frac{1}{(\phi^k)^{(\vert \Lambda_k\vert +\mathcal{M}+1)/2}} 
  \int \exp\left\{\vphantom{\sum_{i\neq m}}-\frac{1}{2}(\eta^k-\zeta^k)^T(\phi^k \Sigma_{\eta^k})^{-1}(\eta^k-\zeta^k)\right.\\
  &\hspace*{0.3cm}\left.-\frac{1}{2\phi^k} (x^k-(1,(x^{\Lambda_k})^T) \eta^k)^2-\frac{1}{2\phi^k}\sum_{i\neq m} (x^{(i),k}-(1,(x^{(i),\Lambda_k})^T) \eta^k)^2\right\} d\eta^k.
\end{aligned}
\end{equation}
We note that the exponent in the integrand is a second order function in $\eta^k$, so by completing the square it is straightforward
to evaluate the integral analytically. When having evaluated the integral and inserting for $f(\phi^k)$ this gives, using
the notation defined in \eqref{gamma_k} to \eqref{Theta_k},
\begin{align*}
    f(\phi^k\vert z^{(m)},x) &\propto \frac{e^{-\frac{1}{\phi^k \beta^k}}}{(\phi^k)^{\alpha^k+1}} \cdot 
  \frac{1}{(\phi^k)^{\mathcal{M}/2}}e^{-\frac{1}{2\phi^k}(\gamma^{m,k}-(\rho^{m,k})^T (\Theta^{m,k})^{-1}\rho^{m,k})} \\
  &= \frac{e^{-\frac{1}{\phi^k \tilde{\beta}^{m,k}}}}{(\phi^k)^{\tilde{\alpha}^k+1}},
\end{align*}
where $\tilde{\alpha}^k$ and $\tilde{\beta}^{m,k}$ are as given in \eqref{alpha_phi_posterior} and \eqref{beta_phi_posterior},
respectively. We recognise this as the density of a inverse gamma distribution, so 
\begin{equation}
    \phi^k\vert z^{(m)},x \sim \text{InvGam}(\tilde{\alpha}^k, \tilde{\beta}^{m,k}).
\end{equation}
This concludes the derivation of the posterior distribution for $\phi$, and we proceed to
the derivation of the posterior for $\eta\vert \phi$. Using the definitions of $f(x\vert  \eta, \phi)$ and $f(\eta\vert \phi)$ we get
\begin{align*}
  f(\eta\vert x, z^{(m)},\phi) &\propto f(\eta\vert \phi)f(x, z^{(m)}\vert \eta, \phi) 
  = f(\eta\vert \phi)f(x\vert \eta, \phi)\prod_{i\neq m}
  f( x^{(m)}\vert \eta, \phi) \\
  &= \prod_{k=1}^K \left[ f(\eta^k\vert \phi^k)f(x^k\vert \eta^k, \phi^k, x^{\Lambda_k})\prod_{i\neq m}
    f( x^{(i),k}\vert \eta^k, \phi^k, x^{\Lambda_k})\right]\\
  &\propto \prod_{i\neq m} f(\eta^k\vert x,z^{(m)},\phi^k),
\end{align*}
where
\begin{align*}
  f(\eta^k&\vert x,z^{(m)},\phi^k) \propto f(\eta^k\vert \phi^k)f(x^k\vert \eta^k, \phi^k, x^{\Lambda_k})\prod_{i\neq m}
    f( x^{(i),k}\vert \eta^k, \phi^k, x^{\Lambda_k})\\
    &\propto \frac{1}{(\phi^k)^{(\vert \Lambda_k\vert +\mathcal{M}+1)/2}}
    \exp\left\{-\frac{1}{2}(\eta^k-\zeta^k)^T(\phi^k \Sigma_{\eta^k})^{-1}(\eta^k-\zeta^k)\right\}\\
    &\times \exp\left\{-\frac{1}{2\phi^k}(x^k-(1,(x^{\Lambda_k})^T) \eta^k)^2-\frac{1}{2\phi^k}\sum_{i\neq m} (x^{(i),k}-(1,(x^{(i),\Lambda_k})^T) \eta^k)^2\right\}.
\end{align*}    
Again introducing $\gamma^{i,k}$, $\rho^{i,k}$ and $\Theta^{i,k}$, as defined in \eqref{gamma_k} to \eqref{Theta_k} this can
be simplified to 
\begin{align*}
    & f(\eta^k\vert x, z^{(m)},\phi^k) & \propto 
  \exp\left\{-\frac{1}{2\phi^k}(\eta^k -(\Theta^{m,k})^{-1}\rho^{m,k})^T\Theta^{m,k}(\eta^k -(\Theta^{m,k})^{-1}\rho^{m,k})\right\},
\end{align*}
which we recognise as the density of a normal distribution, so 
\begin{equation}
    \eta^k \vert x, z^{(m)},\phi^k \sim N((\Theta^{m,k})^{-1}\rho^{m,k}, \phi^k (\Theta^{m,k})^{-1}).
\end{equation}
Thereby the derivation of $f(\eta, \phi\vert x, z^{(m)})$ is complete.

	\bibliography{sn-bibliography}

	
	
	
\end{document}